\documentclass[usegraphicx,usenatbib]{mn2e}

\newcommand{\Halpha}{H{$\alpha$}}

\newcommand{\Hzero}{\ensuremath{H_{\mathrm{0}}}}

\newcommand{\LB}{\ensuremath{L_{\mathrm{B}}}}

\newcommand{\Ldiff}{\ensuremath{L_{\mathrm{diff}}}}
\newcommand{\LFIR}{\ensuremath{L_{\mathrm{FIR}}}}

\newcommand{\LK}{\ensuremath{L_{\mathrm{K}}}}
\newcommand{\Lsol}{\ensuremath{\mathrm{L}_{\odot}}}
\newcommand{\Lsrc}{\ensuremath{L_{\mathrm{src}}}}

\newcommand{\LX}{\ensuremath{L_{\mathrm{X}}}}

\newcommand{\Msol}{\ensuremath{\mathrm{M}_{\odot}}}

\newcommand{\NH}{\ensuremath{N_{\mathrm{H}}}}

\newcommand{\rSN}{\ensuremath{r_{\mathrm{SN}}}}

\newcommand{\Zsol}{\ensuremath{\mathrm{Z}_{\odot}}}

\newcommand{\arcs}{\ensuremath{^{\prime\prime}}}
\newcommand{\degree}{\ensuremath{^\circ}}

\newcommand{\erg}{\ensuremath{\mbox{erg}}}

\newcommand{\nm}{\ensuremath{\mbox{\nm}}}

\newcommand{\ps}{\ensuremath{\s^{-1}}}
\newcommand{\pyr}{\ensuremath{\yr^{-1}}}
\newcommand{\s}{\ensuremath{\mbox{s}}}
\newcommand{\yr}{\ensuremath{\mbox{yr}}}

\newcommand{\ergps}{\ensuremath{\erg~\ps}}

\newcommand{\Msolpy}{\ensuremath{\Msol~\pyr}}

\newcommand{\CHANDRA}{\emph{Chandra}}

\newcommand{\ROSAT}{\emph{ROSAT}}

\newcommand{\XMM}{\emph{XMM-Newton}}

\newcommand{\chisq}{\ensuremath{\chi^2}}

\title{The {\em Chandra} View of Galaxy Mergers}

\author[Brassington et al.]
  {Nicola J. Brassington,$^1$\thanks{E-mail: njb@star.sr.bham.ac.uk} Trevor J. Ponman$^1$ and Andrew M. Read$^2$ \\
  $^1$School of Physics and Astronomy, The University of Birmingham,
  Edgbaston, Birmingham B15 2TT, UK\\
  $^2$Department of Physics and Astronomy, University of Leicester,University Road, Leicester LE1 7RH, UK}

\date{Accepted 0000 00. Received 0000 00; in original form 0000 00}

\begin{document}

\maketitle

\begin{abstract}

From a {\em Chandra} survey of nine interacting galaxy systems the evolution of
X-ray emission during the merger process has been investigated. It is found that the X-ray luminosity peaks $\sim$300 Myr before nuclear coalescence, and then dips, even though we know that rapid and increasing activity is still taking place at this time. It is likely that this drop in X-ray luminosity is a consequence of outflows breaking out of the galactic discs of these systems. In this work it is also shown that, for the systems close to the point of nuclear coalescence, \LFIR\ becomes massively enhanced compared to the X-ray luminosity of these systems. We suggest that this enhancement may indicate a `top heavy' IMF, with  an enhanced fraction of massive stars.

At a time $\sim$1 Gyr after coalescence, the merger-remnants in our sample are X-ray faint when compared to typical mature elliptical galaxies. However, we do see evidence that these systems will start to resemble typical elliptical galaxies at a greater dynamical age, given the properties of the 3 Gyr system within our sample, supporting the idea that halo regeneration will take place within low {\ensuremath{L_{\mathrm{X}}}} merger-remnants.

As a part of this survey, detailed \CHANDRA\ observations for the double nucleus merger system Mkn 266 and the merger-remnant Arp 222 are presented for the first time. With the Mkn 266 observation, in contrast to previous studies, we now have good spectral information of the individual components part-seen with the \ROSAT\ HRI. Additionally, the structure of the emission to the north of the system can clearly be distinguished and there is also a suggestion of some extension of X-ray emission to the south east of the nuclear region, indicating that this galaxy could just be on the verge of large-scale galactic winds breaking out. Within Arp 222 an X-ray luminosity of 1.46 $\times$10$^{40}$erg s$^{-1}$ has been detected, this is the lowest value of \LX\ within our sample. The diffuse gas of Arp 222 has been modelled with a temperature of 0.6 keV and, from CO observations it has been found to host very little molecular gas, indicating that, from current observations, Arp 222 does not resemble a mature elliptical.

\end{abstract}


\section{Introduction}

It is widely believed that very few galaxies exist today that have not been formed or shaped in some way by an interaction with another galaxy. \citet{Toomre_77} identified 11 galaxies that exhibit characteristics of on-going mergers, and arranged them in a chronological order, illustrating how spiral galaxies can merge to produce ellipticals. It is now the view of many astronomers that this process, whilst not the only mechanism to create these systems, plays a vital role in the production of elliptical galaxies. The 11 systems within the `Toomre' sequence, alongside other examples of on-going mergers, have been studied in great detail over a range of wavelengths to try and characterise exactly how these systems evolve. When studying these merging galaxies, X-ray observations are of particular importance, as they are able to probe the dusty nucleus of the system, which can be obscured at other wavelengths, allowing the nature of the point source population to be established. Also imaging of the soft X-ray emission permits the diffuse hot gas to be mapped out. This gives important information about the hot gaseous component associated with the strong starburst, and  allows galactic-winds outflowing from the system to be observed, enabling constraints to be placed on the energetics of these outflows.

X-ray observations were initially carried out with the {\em Einstein Observatory}, providing limited spatial resolution. This instrument was followed by \ROSAT, which greatly improved the sensitivity of the observations, allowing the X-ray properties of these merging galaxies to be probed. A study of a sample of interacting systems was carried out by \citet{Read_98} (from here on RP98), where the X-ray luminosity and properties of the diffuse gas were investigated with \ROSAT, alongside the point source population. In this study it was found that the normalised X-ray luminosity of these systems, broadly speaking, followed the normalised \LFIR\ luminosity, peaking at the time of coalescence. It was also found that the young merger-remnant systems within the sample exhibited low X-ray luminosities, with no indication from the later stage systems in their sample that these systems would increase in X-ray luminosity. However, it was noted that the $\sim$1 Gyr system, NGC 7252, still hosts a large amount of molecular gas, in the form of tails and loops, and given more time to evolve, the X-ray properties could resemble those of a mature elliptical galaxy at a greater dynamical time. The main limitation of this study, due to the \ROSAT\ observations, was the inability to disentangle the point sources from the diffuse gas, particularly at larger distances. 

With the next generation of X-ray observatories this issue was addressed with an increase in spatial resolution. This improvement, provided by \CHANDRA, has allowed the investigation of both the diffuse gas and the point source population of galaxies to be carried out in greater detail. The ability to disentangle these two components is vital when investigating interacting and merging galaxies, as studies have shown that both of these components have different evolutionary timescales \citep{Read_01}. It is therefore important to be able to study both of these separately, to fully understand how they evolve.

From observing a selection of interacting galaxy systems, at different stages of evolution, the processes involved in the merging of galaxies pairs can be characterised. In the following sections we will describe the sample we have selected, and investigate the behaviour of the X-ray emission as the galaxies evolve from two spiral galaxies, through to relaxed merger-remnants. Section \ref{sec:sample} outlines the selection criteria we have used and gives a brief description of each system. In section \ref{sec_data_red} we present the results from  new \CHANDRA\ observations of Mkn 266 and Arp 222. Correlations and evolution of X-ray emission across the whole sample is presented in section \ref{sec_evol}, and discussed in section \ref{sec:dis}. Conclusions are given in section \ref{sec:con}.

\section{The Sample}
\label {sec:sample}

To gain a better understanding of galaxy evolution it is important to compare similar systems that are undergoing the same transformation. By compiling a sample of these galaxies, the evolution of the systems' X-ray properties can be investigated. In this paper we have a sample of nine interacting and post-merger systems, carefully selected to ensure that they are representative of galaxies undergoing a major merger. The nine systems were selected using the following criteria;

\begin{enumerate}
\item 
All systems have been observed with {\em Chandra}.
\item 
Systems comprise of, or, originate from, two similar mass, gas rich, spiral galaxies.
\item
Multi-wavelength information is available for each system.
\item
The absorbing column is low, maximising the sensitivity to soft X-ray emission.
\item
A wide chronological sequence is covered; from detached pairs to merger-remnants.
\end{enumerate}

Once the nine systems to study had been selected, the issue of merger age had to be addressed. This is one of the main problems when working with a chronological study such as this, and a number of different methods are required to solve this problem. Firstly the point of nuclear coalescence was assigned to be at time 0. From this, the time taken until nuclear coalescence for each pre-merger system can then be estimated. These timescales were derived with a combination of N-body simulations, such as \citet{Mihos_96}, and dynamical age estimates, where the length and faintness of tidal tails, as well as nuclei separation, were used \citep{Toomre_72}. For post-merger systems, assigning an age estimate was done by making the assumption that the last widespread episode of star formation within the system took place at the time of nuclear coalescence. Stellar population synthesis models were then used to calculate these timescales, therefore giving a good merger age estimate \citep{Bruzual_93}. In the following sub-sections a brief description of the nine merger examples, and their X-ray properties, are presented in chronological order.

\subsection{Arp 270}

The earliest example of an interacting system in our sample is Arp 270 (also NGC 3395/3396). This comprises two spiral galaxies of comparable mass \citep{Hern_01}, separated by 12\,kpc at a distance of 28 Mpc (assuming \Hzero\ = 75\,km s$^{-1}$
Mpc$^{-1}$, and accounting for Virgocentric in-fall). These galaxies are connected by an optical bridge which is thought to have formed during the systems first perigalactic passage which occurred approximately 5$\times$10$^8$ years ago.
 
\begin{figure}
  \includegraphics[width=\linewidth]{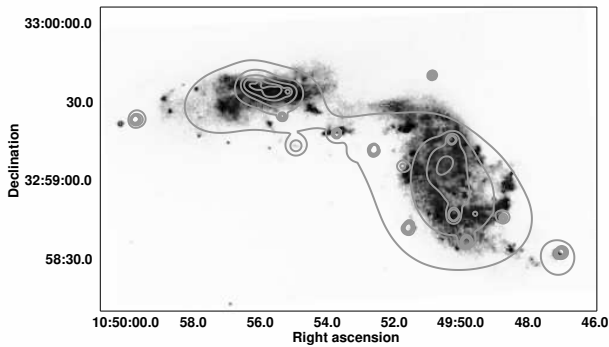}
  \hspace{0.1cm}
  \caption{The early merger system Arp 270. Contours of adaptively smoothed 0.3$-$8.0\,keV X-ray data from {\em Chandra} ACIS-S overlaid on an optical image from the Palomar 5m telescope. }
  \label{fig:arp270_optic_con}
\end{figure}

A 20 ks \CHANDRA\ observation of the system was made in 2001 and is discussed in detail in \citet{Brassington_05}, the contours of adaptively smoothed 0.3$-$8.0\,keV X-ray contours overlaid on an optical image are shown in Figure \ref{fig:arp270_optic_con}. From this observation 16 point sources are detected, 7 of which are classified as {\em Ultraluminous X-ray Sources} (ULX's), with \LX\ $\ge 1\times$10$^{39}$erg s$^{-1}$ \citep{Soria_05b}. The diffuse gas emits at a global temperature of $\sim$0.5 keV and shows no evidence of hot gaseous outflows as are seen in later stage systems (RP98). The galaxy pair, although in a very early stage of interaction, already show increased levels of \LFIR\ compared to quiescent galaxies, indicating that there is enhanced star formation taking place. The numerical simulations of \citet{Mihos_96} suggest that a system such as Arp 270 will coalesce in $\sim$650 Myrs.

\subsection{The Mice}

\begin{figure}
  \includegraphics[width=\linewidth]{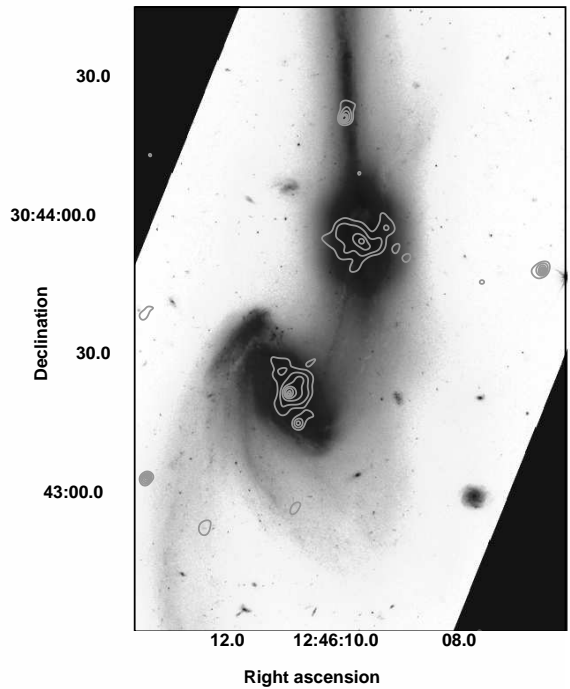}
  \hspace{0.1cm}
  \caption{The Mice, contours of adaptively smoothed 0.2$-$10.0\,keV X-ray data from {\em Chandra} ACIS-S overlaid on an optical image from WFPC2 on board the {\em HST}. }
  \label{fig:mice_opt_con}
\end{figure}

The Mice (also Arp 242, NGC  4676A/B) is another early stage merger system, lying second in the evolutionary sequence proposed by \citet{Toomre_72}. At a distance of 88 Mpc, the two detached spiral galaxies are again connected by a tidal bridge and, in addition to this feature, exhibit large tidal tails, a consequence of the galaxies interacting as they have passed each other. \citet{Read_03} reports on a 30.5\,ks \CHANDRA\ observation of the system, the adaptively smoothed 0.2$-$10.0\,keV X-ray contours overlaid on an optical image are shown in Figure \ref{fig:mice_opt_con}. 

Five ULXs are detected in association with the galaxies, these sources have been found to be coincident with regions of ongoing star formation, both within the nuclei of the galaxies and also in the tidal tails. The structure of the diffuse X-ray gas in both nuclei suggests that this system is at a more advanced stage of evolution than Arp 270. This is indicated by the morphology of the gas, a soft, thermal plasma, which extends out of the minor axis of both galaxies, suggesting that these features are starburst driven winds. As in Arp 270, the Mice emits an enhanced level of \LFIR, again indicating that this system has enhanced star formation taking place, this emission is particularly high in the nuclei of the galaxies. 

\subsection{The Antennae}

\begin{figure}
  \includegraphics[width=\linewidth]{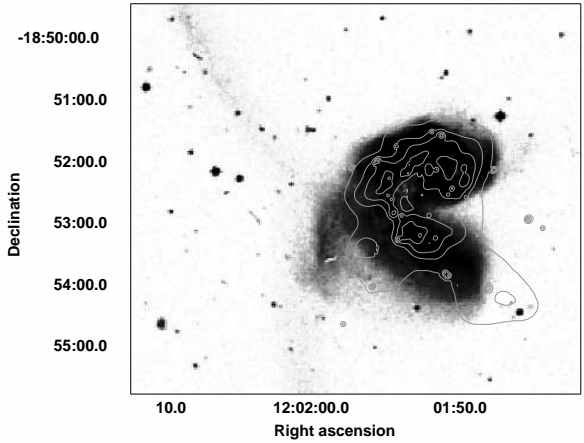}
  \hspace{0.1cm}
  \caption{The Antennae, contours of adaptively smoothed 0.3$-$6.0\,keV X-ray data from {\em Chandra} ACIS-S overlaid on an optical image from the UK Schmidt telescope. }
  \label{fig:ant_opt_con}
\end{figure}

The Antennae, NGC 4038/4039, or Arp 244, is probably the most famous example of a galaxy pair undergoing a major merger, and, lying at a distance of 19 Mpc, is also the nearest. A deep integrated \CHANDRA\ observation of 411\,ks has been made of this system, enabling the nature of both the point source population and the diffuse gas to be investigated \citep{Fabbiano_04, Zezas_04}, Fig \ref{fig:ant_opt_con} shows the full band (0.3$-$6.0 keV) X-ray contours overlaid on an optical image. 

Due to the depth of the observation, sources were detected down to a luminosity of 2$-$5 $\times$10$^{37}$erg s$^{-1}$. This resulted in a detection of 120 point sources, 12 of these have been confirmed as ULXs. The integrated observation comprises 7 separate pointings, enabling the variability of the point sources to be investigated.

Out of the 12 ULXs, 4 emit below 1 $\times$10$^{39}$ erg s$^{-1}$, the lower luminosity threshold for a source to be classified as a ULX, in at least one observation. Further, one of the sources was observed in only one pointing, indicating that it is likely to be a transient, providing further evidence that ULXs are a heterogeneous class, comprising of contributions from X-ray binary systems, transient sources and also, possibly, intermediate mass black holes (IMBHs). 

The quality of the data has also lead to detailed mapping of the diffuse gas, investigating both the temperature and metallicity variation of the ISM. 21 separate regions of the diffuse gas have been analysed, spectral fitting of these regions reveal that there is a variation in temperature of the diffuse gas from 0.2 to 0.9 keV and metallicities vary from regions of sub-solar abundances to areas of emission displaying super-solar abundances, notably in both the nuclear regions and two hotspots in the northern loop of the disc (R1 and R2 reported in \citet{Fabbiano_03b}). 

The morphology of the gas reveals that there are large scale diffuse features; two large faint X-ray loops extending to the South of the system and a low-surface-brightness halo in the region surrounding the stellar discs extending out to $\sim$18\,kpc from the nucleus of NGC 4039. The two loops have temperatures ranging from 0.29 to 0.34 keV and the low surface brightness halo has a nominal temperature of 0.23 keV. This cooler larger scale emission may be the aftermath of a superwind, possibly from the first encounter of the two systems, which took place $\sim$2$-$5 $\times$10$^8$ years ago. The star formation rates within the two nuclei are 2.1 \Msolpy\ and 1.7 \Msolpy\ and in the region where the discs overlap has been found to be 5.0 \Msolpy\ \citep{Mihos_93}.


\subsection{Mkn 266}

The next system within this sample, Markarian 266 (also NGC 5256), is a double nucleus system with a large gaseous envelope. Here, a summary of the galaxy properties are presented, with the full analysis and results of a recent \CHANDRA\ observation detailed in section \ref{sec:mkn266}. 

\begin{figure}
  \includegraphics[width=\linewidth]{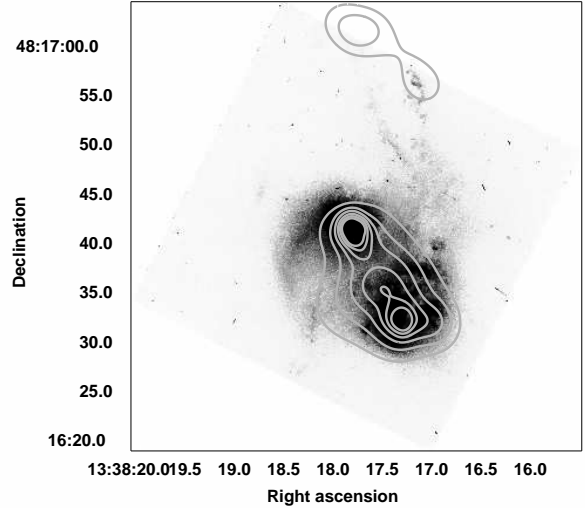}
  \hspace{0.1cm}
  \caption{Mkn 266, contours of adaptively smoothed 0.3$-$8.0\,keV X-ray data from {\em Chandra} ACIS-S overlaid on an optical image from WFPC2 on board the {\em HST}. }
  \label{fig:mkn_266_opt_con}
\end{figure}

This system lies at a distance of 115 Mpc and has an X-ray luminosity of 7.32 $\times$10$^{41}$erg s$^{-1}$. The two nuclei, originating from the progenitors, are thought to comprise a LINER with a powerful starburst component, to the north of the system, and a Seyfert type 2 to the South \citep{Mazzarella_88}. In addition to these sources, an area of enhanced emission has been detected between these nuclei, it is thought that this is caused by the collision of the two discs. Prior to a 20\,ks \CHANDRA\ observation of Mkn 266, made in 2001, this feature had not been seen in X-rays before. But, due to {\em Chandra's} superior spatial resolution, it is now possible to distinguish this feature. This can be seen in Figure \ref{fig:mkn_266_opt_con}, where the adaptively smoothed, full band X-ray contours (0.3$-$8.0\,keV) overlaid on an image from WFPC2, on board The {\em Hubble Space Telescope (HST)} are shown.

In addition to the central emission from the colliding galaxies, a large X-ray gas cloud can be seen to the North of the system. From a \ROSAT\ observation it was proposed that this feature is a superwind, driven by the effect of the starburst's supernovae and stellar winds from the centre of the system \citep{Wang_97}. \citet{Kollatschny_98} argue that an X-ray `jet', arising from a centrally powered superwind, is unlikely due to both the non-radial geometry of the emission and also the bright X-ray emission of the feature. They suggest that the most plausible mechanism to explain the `jet' is from excitation by hot post-shock gas, although they do not conclude where the energy to power this feature would arise from.  From the \CHANDRA\ observation we speculate that the northern emission observed could also arise from star formation taking place in a tidal arm that has been stripped from the galaxy during an earlier interaction. The exact nature of this emission is discussed in detail in section \ref{sec:mkn266_northern}.

\subsection{NGC 3256}

\begin{figure}
  \includegraphics[width=\linewidth]{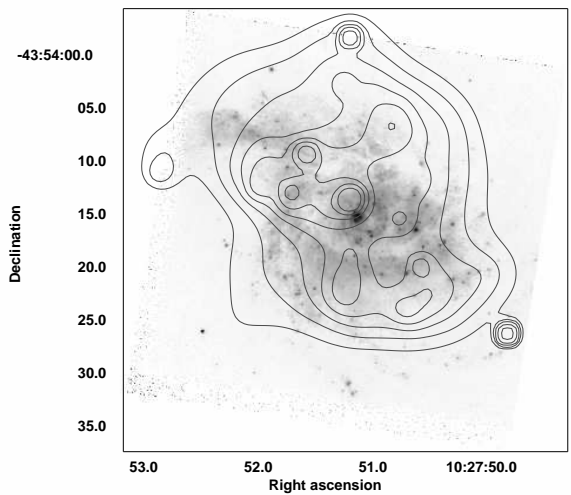}
  \hspace{0.1cm}
  \caption{NGC 3256, contours of adaptively smoothed 0.3$-$8.0\,keV X-ray data from {\em Chandra} ACIS-S overlaid on an optical image from WFPC2 on board the {\em HST}. }
  \label{fig:ngc3256_opt_con}
\end{figure}

The most X-ray luminous system in our sample, NGC 3256, is a powerful ultraluminous infrared galaxy (ULIRG), lying at a distance of 56 Mpc. This merger system, like Mkn 266, has one common gaseous envelope containing the nuclei from its two parent galaxies. \citet{Lira_02} report on a 28\,ks \CHANDRA\ observation made in 2000. Contours of adaptively smoothed 0.2$-$8.0\,keV X-ray emission overlaid on an optical image of the system are shown in Figure \ref{fig:ngc3256_opt_con}. 

The total X-ray luminosity of NGC 3256 is 7.87$\times$ 10$^{41}$ \ergps\ with $\sim$70\% of the X-ray luminosity arising from the diffuse emission. 14 discrete X-ray sources were detected in this system, all of which have been classified as ULXs. Due to the high source detection threshold of this observation (\LX=1.4$\times$10$^{39}$erg s$^{-1}$) fainter point sources were not detected. Both galaxy nuclei are clearly detected in X-rays. The Northern nucleus is a site of intense star formation, and UV spectra \citep{Lipari_00} show strong absorption lines, implying the presence of massive young stars. The Southern nucleus is heavily obscured and appears to be less active then the Northern nucleus. It has been suggested that it hosts an AGN, although there is no clear evidence of this provided in the X-ray data. The soft diffuse emission of the system can be described by two thermal components with a harder tail. The thermal plasma components exhibit temperatures of 0.6 keV and 0.9 keV and the hard component is thought to arise from a contribution from the lower luminosity X-ray point source population. 

A kinematic study of the system \citep{English_03} suggests that NGC 3256 is currently experiencing the starburst that just precedes the final core collision and, given the close proximity of the two nuclei, it is likely that this system has undergone more than one perigalactic approach. Assuming that the two tidal tails formed during the last closest-encounter, the time that has elapsed since then can be estimated. This characteristic timescale was calculated to be 500 Myr and it is thought that coalescence will take place in $\sim$200 Myr.

\subsection{Arp 220}

\begin{figure}
  \includegraphics[width=\linewidth]{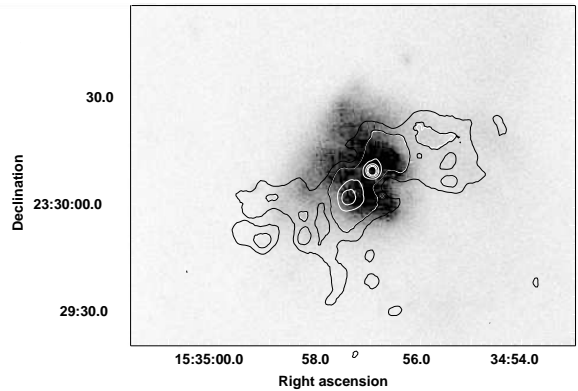}
  \hspace{0.1cm}
  \caption{Arp 220, contours of adaptively smoothed 0.2$-$1.0\,keV X-ray data from {\em Chandra} ACIS-S overlaid on an optical image from the Palomar 5m telescope. }
  \label{fig:arp220_opt_con}
\end{figure}

The double nuclear system, Arp 220 (IC 4553/4, UGC 09913) is a prototypical ULIRG, lying at a distance of 76 Mpc. The two nuclei have a separation of less than 0.5 kpc and are just at the point of coalescence. This complicated, dusty system has the highest value of \LFIR\ in our sample, which arises from the large reservoir of hot gas within the system. The mechanism by which this gas is heated is still currently under debate. Both the presence of a heavily shrouded AGN and intense nuclear starbursts have been suggested, along with a combination of both these energy sources contributing to the heating mechanism. A 60 ks \CHANDRA\ observation of the system was made in 2000, and from these data both the nuclear and the extended emission have been analysed \citep{Clements_02,McDowell_03}. Figure \ref{fig:arp220_opt_con} shows the 0.2$-$1.0 keV adaptively smoothed X-ray emission, overlaid on an optical image.

The system hosts 4 ULXs, including the nuclei, no other point sources are detected. Three distinct regimes of large diffuse gas structures are observed; the circumnuclear region, a plume region and, further out, diffuse lobe regions. The circumnuclear region has both compact, hard nuclear emission that resides in the 1 kpc at the centre of the system and softer, more extended emission. It has been suggested that the hard component of this region arises from a significant point source population, emitting at a luminosity lower than $\sim 5 \times 10 ^{39}$ \ergps, the sensitivity threshold for this observation. Although, this could also be a AGN, albeit one emitting at a low luminosity.

The plume regions extend to the northwest and southeast of the system with a projected tip to tip length of $\sim$10 kpc, these features are clearly seen in Figure \ref{fig:arp220_opt_con}. The spectrum of the plumes include both hot (1$-$5 keV) and cooler (0.25 keV) thermal contributions. It is likely that these regions are associated with a superwind extending from the nuclear region as a consequence of the vigorous star formation that is taking place at the centre of the system. Beyond the plumes, two large, low surface brightness lobe regions have been observed extending from 10$-$15 kpc on either side of the nuclear region. These regions have been found to be cooler than the plumes (0.2$-$1.0 keV), although, due to the low number of counts, higher temperature gas residing in these regions cannot be ruled out. 

It was noted by \citet{Heckman_96} that the plumes and lobes are ``misaligned'' by 25\degree$-$30\degree, they suggest that this could be due to a change in orientation of the system as the encounter has progressed. \citet{McDowell_03} propose that this misalignment is actually a consequence of the lobes being produced not by the superwinds in the system, but are a product of the merger itself. Tentative evidence to support this scenario is presented in an INTEGRAL observation by \citet{Colina_04}.

\subsection{NGC 7252}

NGC 7252, the first example of a merger-remnant galaxy in our sample, is a proto-typical merger galaxy at a distance of 63 Mpc. The central part of the system, a single relaxed body, displays an r$^{1/4}$ optical surface brightness profile typical of elliptical galaxies \citep{Schweizer_82}. In addition to this relaxed nucleus, the galaxy exhibits complex loops and ripples, notably two long tidal tails, indicative of the merger history of this system.  From both {\em UBVI} images taken with the WCFPT instrument on the {\em HST} \citep{Miller_97} and N-body simulations \citep{Hibbard_95}, it has been estimated that nuclear coalescence took place $\sim$1 Gyr ago.

\begin{figure}
  \includegraphics[width=\linewidth]{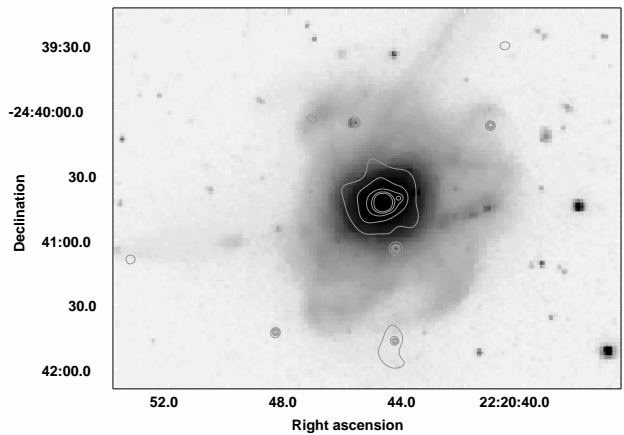}
  \hspace{0.1cm}
  \caption{NGC 7252, contours of adaptively smoothed 0.3$-$7.0\,keV X-ray data from {\em Chandra} ACIS-S overlaid on an optical image from the LCO 2.5m telescope. }
  \label{fig:ngc_7252_opt_con}
\end{figure}

A 28 ks \CHANDRA\ observation, along with an \XMM\ observation is reported in \citet{Nolan_04}. Figure \ref{fig:ngc_7252_opt_con} shows the 0.3$-$7.0 keV adaptively smoothed \CHANDRA\ X-ray contours overlaid on an optical image. A total of nine ULXs are detected within the optical confines of this system, a further 5 sources are also detected but at a lower significance ($<3\sigma$). The hot, diffuse gas of the system has been found to be fairly symmetrical and has an X-ray luminosity of 2.42$\times$10$^{40}$ \ergps. This is low when compared to luminosities from typical elliptical galaxies (\LX\ $\sim$10$^{41-42}$ \ergps). The low luminosity of this system is possibly due to the young age of the galaxy and, over much longer timescales, X-ray halo regeneration may increase the mass and hence luminosity of the gas within this system to levels seen in typical elliptical galaxies. From spectral modelling of the \XMM\ data the X-ray emission from the nuclear region is found to emit at 0.72 keV and 0.36 keV. There is also a harder contribution that can be fitted with a power law, and it is expected that this is due to a lower luminosity point source population. 

During the nuclear coalescence of NGC 7252 the star formation rate at the centre of the galaxy would have been massively enhanced. Now, as the reserve of gas becomes depleted, the star formation rate has fallen to one third of the value it would have been at its peak \citep{Mihos_93}, although its \LFIR\ is still enhanced when compared to normal quiescent galaxies.

\subsection{Arp 222}

Arp 222, or NGC 7727, is a post merger galaxy at a slightly more advanced stage of evolution than NGC 7252 \citep{Georgakakis_00}. A CO observation \citep{Crabtree_94} indicates that these systems are very similar in morphology and both host young cluster populations, however, Arp 222 seems to have a much smaller molecular gas content than NGC 7252. From a K-band observation \citep{Rothberg_04} it has been found that the optical surface brightness profile of Arp 222 follows the de Vaucouleurs r$^{1/4}$ law, indicating that violent relaxation has taken place since the merger. From analysing the discrete structure of the galaxy, plumes extending from the nucleus were identified, these features indicate that the system has still not fully relaxed into a mature elliptical galaxy.

\begin{figure}
  \includegraphics[width=\linewidth]{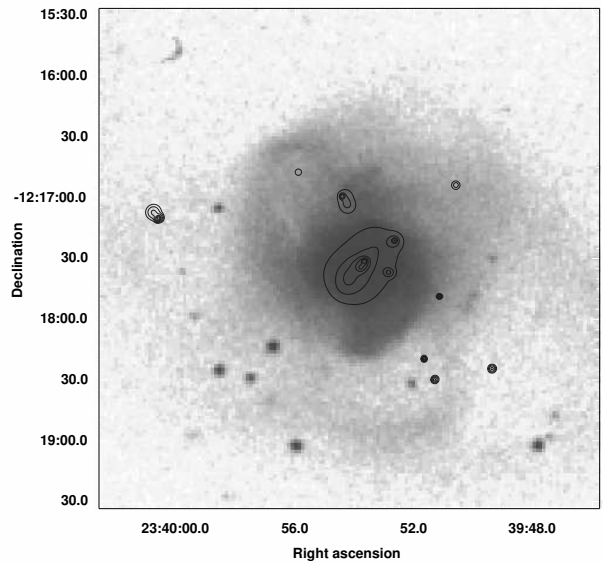}
  \hspace{0.1cm}
  \caption{Arp 222 Contours of adaptively smoothed 0.3$-$8.0\,keV X-ray data from {\em Chandra} ACIS-S overlaid on an optical image from the UK Schmidt telescope. }
  \label{fig:arp222_opt_con}
\end{figure}

A 19 ks \CHANDRA\ observation of the post merger-remnant was carried out in 2001, and analysis and results from this observation are given in section \ref{sec:arp222}. Figure \ref{fig:arp222_opt_con} shows the 0.3$-$8.0 keV adaptively smoothed X-ray emission overlaid on an optical image. From this observation 15 point sources were detected, two of which are classified as ULXs. The X-ray luminosity of the diffuse gas in this system is 6.52 $\times$10$^{39}$\ergps, much lower than the X-ray luminosity emitted from NGC 7252. This is likely to be due to the smaller gas content of Arp 222, which, as can be seen in Figure \ref{fig:arp222_opt_con}, does not extend to the optical confines of the galaxy. From spectral modelling the global temperature of the diffuse gas in this system has been found to be 0.60 keV.
The \LFIR\ value of the system is much lower than that of the previous systems and is similar to values seen in elliptical galaxies. 

\subsection{NGC 1700}

\begin{figure}
  \includegraphics[width=\linewidth]{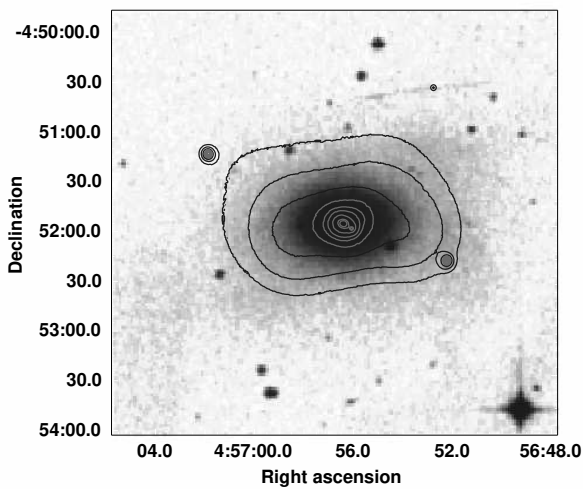}
  \hspace{0.1cm}
  \caption{NGC 1700 Contours of adaptively smoothed 0.3$-$0.8\,keV X-ray data from {\em Chandra} ACIS-S overlaid on an optical image from the UK Schmidt telescope. }
  \label{fig:ngc1700_opt_con}
\end{figure}

The final system in our sample, NGC 1700, is a protoelliptical galaxy with a best age estimate of $\sim$3 Gyr \citep{Brown_00}. This galaxy possesses a kinematically distinct core and boxy isophotes at larger radii, it also contains two symmetrical tidal tails, a good indicator that this system formed through the merger of two, comparable mass, spiral galaxies. It has been suggested that it is the presence of these tidal features that causes the 'boxiness' of the galaxy \citep{Brown_00}. A 42 ks \CHANDRA\ observation was made of the system in 2000, and the results of this are reported in \citet{Statler_02}. Figure \ref{fig:ngc1700_opt_con} shows the 0.3$-$0.8 keV adaptively smoothed X-ray emission overlaid on an optical image of NGC 1700.

From this observation, 36 point sources are detected, 6 of which are classified as ULXs \citep{Diehl_06}. The diffuse X-ray gas is well modelled by a single temperature thermal plasma at a temperature of 0.43 keV and has an X-ray luminosity of 1.47$\times 10^{41}$ \ergps, similar to that of a mature elliptical galaxies. The change in morphology of the diffuse X-ray gas from the central elliptical region to the outer boxy region, can clearly be seen in Figure \ref{fig:ngc1700_opt_con}. \citet{Statler_02} suggest that the flattening of the isophotes is a consequence of an elliptical-spiral interaction, not a spiral-spiral merger as suggested by \citet{Brown_00}. They argue that an interaction involving a preexisting elliptical with a hot ISM, would lead to the channelling of the already hot gas into the systems common potential well. This gas, at sufficiently low densities, would then settle into a rotationally flattened cooling disc, as is observed. Currently, neither these two scenarios, nor the suggestion that the system could have formed from a 3-body interaction \citep{Statler_96}, can be ruled out. 

\section{{\em Chandra} Observations and Data Analysis of Mkn 266 and Arp 222}
\label{sec_data_red}

Observations of Mkn 266 and Arp 222 were carried out with the ACIS-S camera on board the {\em Chandra} X-ray Observatory. Mkn 266 was observed on 2nd November 2001 with a total observation time of 19.7\,ks and Arp 222 was observed on 18th December 2001 with a total observation time of 19.0\,ks. The initial data processing to correct for the motion of the
spacecraft and apply instrument calibration was carried out with the
Standard Data Processing (SDP) at the {\em Chandra} X-ray Center
(CXC). The data products were then analysed using the CXC CIAO
software suite (v3.2)\footnote{http://asc.harvard.edu/ciao} and
HEASOFT (v5.3.1). The data were reprocessed, screened for bad pixels, and time filtered to remove periods of high background (when
counts deviated by more than 5$\sigma$ above the mean). This
resulted in a corrected exposure time of 18.5ks for Mkn 266 and 18.8\,ks for Arp 222. The full data analysis and results from these two observations are detailed in the following subsections.

\subsection{Mkn 266}
\label{sec:mkn266}
\subsubsection{Overall X-ray Structure}
\label{mkn:sec_struc}

\begin{figure*}
  \begin{minipage}{0.5\linewidth}
  \vspace{0.7cm}
  \includegraphics[width=\linewidth]{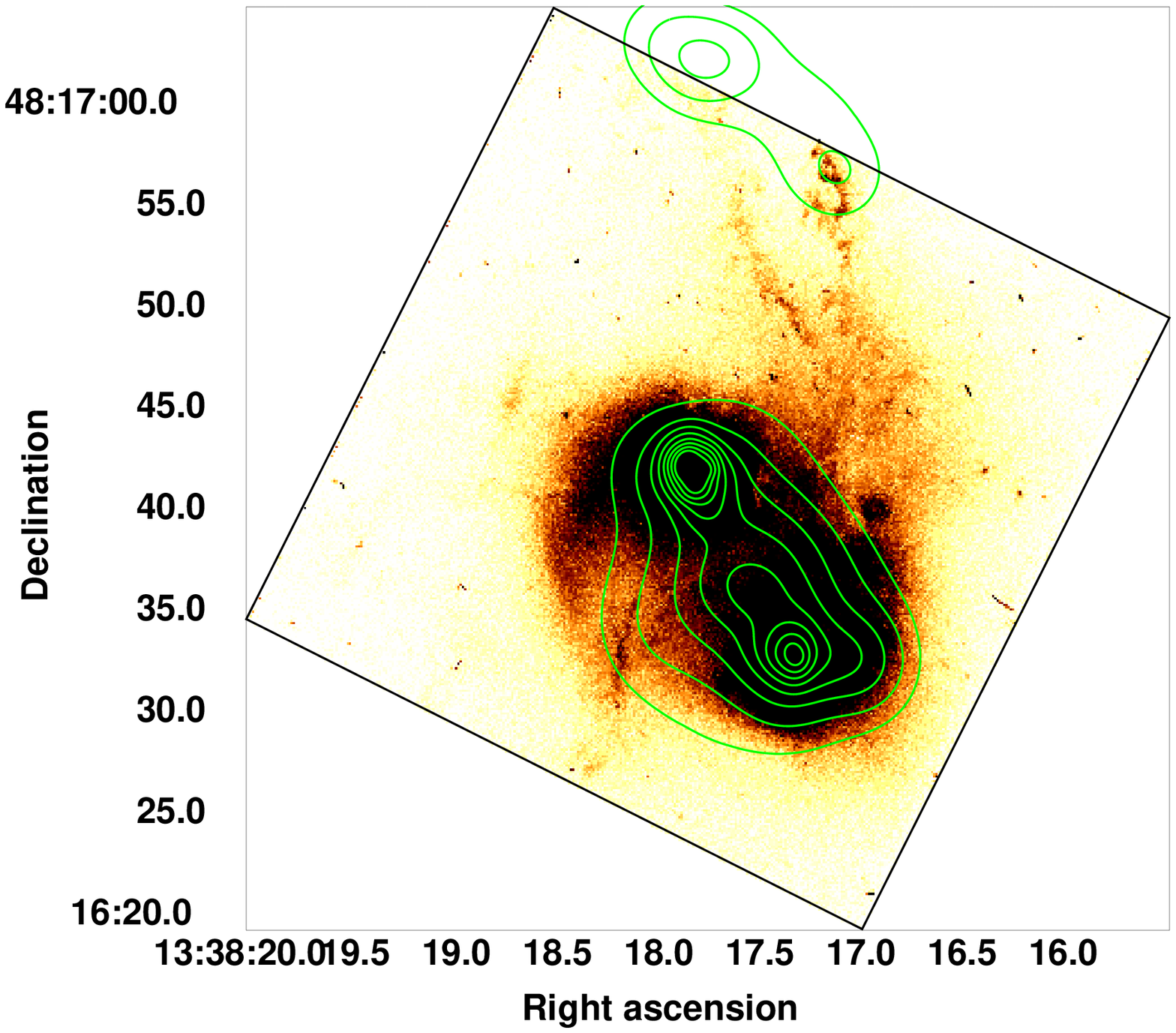}

  \end{minipage}\vspace{0.05\linewidth}
  \begin{minipage}{0.38\linewidth}

  \hspace{0.1\linewidth}
  \includegraphics[width=\linewidth]{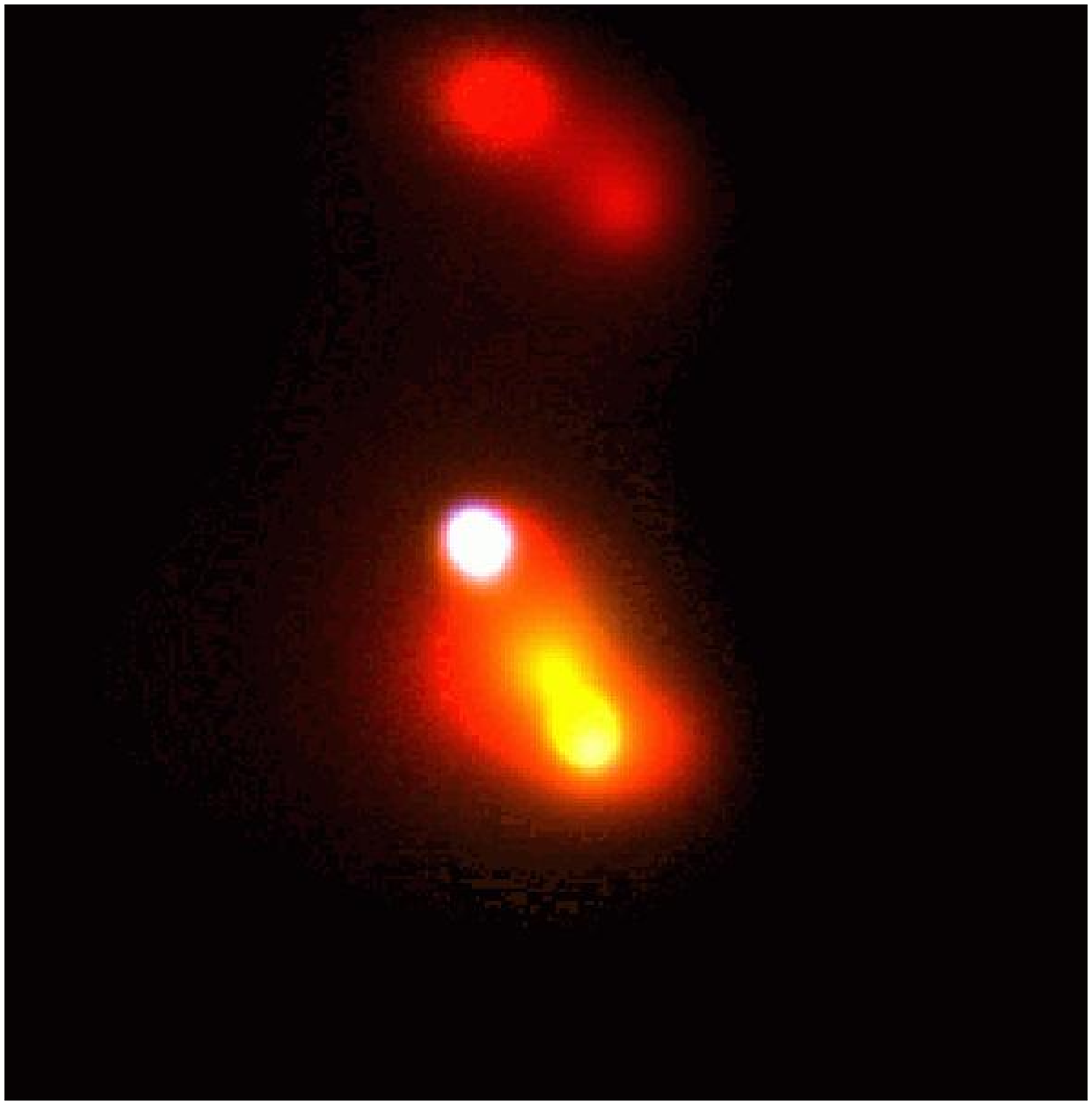}

  \end{minipage}
  \caption{Left, contours of adaptively smoothed 0.3$-$8.0\,keV X-ray data from {\em Chandra} ACIS-S overlaid on an optical image from WFPC2 on board the {\em HST}. The black box indicates the plate boundary of the {\em HST} image. Right shows the `true colour' image of Mkn 266. Red corresponds to 0.3$-$0.8 keV, green to 0.8$-$2.0 keV and blue to 2.0$-$8.0 keV.}
  \label{fig:mkn_rgb}
\end{figure*}

A 0.3$-$8.0\,keV (from here on referred to as `full band') {\em Chandra}
image was created from the cleaned events file and adaptively smoothed
using the CIAO task {\em csmooth} which uses a smoothing kernel to
preserve an approximately constant signal to noise ratio across the
image, which was constrained to be between 2.6 and 4. In Figure
\ref{fig:mkn_rgb} both the optical image from the {\em HST} with the full band X-ray contours overlaid (left) and the `true colour' image of the galaxy system (right) are shown. The `true colour' image was created by combining three separate smoothed images in three energy bands; 0.3$-$0.8 keV, 0.8$-$2.0 keV and 2.0$-$8.0 keV, using the same smoothing scale for each image. These energy bands correspond to red, green and blue respectively. From these images it can be seen that the system is comprised of two separate regions of X-ray emission. To the South of the images the central emission from the interacting galaxies can be seen. This comprises of the two nuclei from the progenitor galaxies, a LINER to the North and a Seyfert 2 to the South, contained within a common gaseous envelope and a region of enhanced emission between the two nuclei. This region is coincident with a radio source reported in \citet{Mazzarella_88}. It is likely that this enhanced emission is due to the interaction of the two galaxy discs. To the North of the images diffuse gas is detected and shows some correlation with emission seen in the {\em HST} image. The nature of this emission is discussed in detail in section \ref{sec:mkn266_northern}.

From the `true colour' image it can be seen that the north-east nucleus emits in all three energy bands, whilst the south-west nuclear emission is not as hard. There is also some enhanced X-ray emission in the central region between the two nuclei. The X-ray emission surrounding these nuclei appears to be soft and diffuse with some suggestion of a super-bubble to the south-east of the system.

\subsubsection{Point Source Spatial and Spectral Analysis}
\label{mkn:sec_source_ps}

\begin{figure}
  \hspace{0.1\linewidth}
  \includegraphics[width=0.8\linewidth]{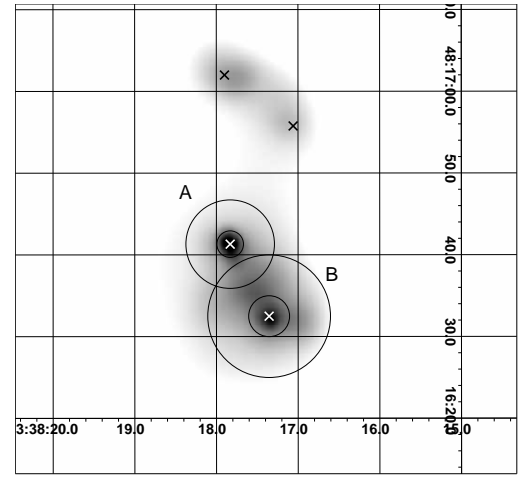}
  \hspace{0.1cm}
  \caption{Adaptively smoothed 0.3$-$8.0\,keV X-ray data from {\em Chandra} ACIS-S of Mkn 266, with {\em wavdetect} point source denoted  with an ``X''. Extraction and background region files for source A and B are also shown. }
  \label{fig:mkn_reg}
\end{figure}

Discrete X-ray sources were detected using the CIAO tool {\em
wavdetect}. This was run on the full band
image, over the 2, 4, 8, 16 pixel wavelet scales (where pixel
width is 0.1\arcs), with a significance threshold of
$2.5\times10^{-5}$, which corresponds to one spurious source over a
200 $\times$ 200 pixel grid, the size of our image. Only 4 sources were
detected in this range; the two nuclei and two regions in the diffuse northern feature. Both the regions in the northern emission contained less than 30 counts and had large detection regions of r$\ge$3.5\arcsec\ and so were not defined as point sources. Instead, a spectrum was extracted from a region file containing all the northern emission. These two detections, along with the extraction and background extraction regions for the 2 nuclei, are shown in Figure \ref{fig:mkn_reg}.

The two detected nuclear sources were extracted using the source
region files created by {\em wavdetect}. The size of each region was
selected to ensure that as many source photons as possible were
detected whilst minimising contamination from nearby sources and
background. The background files were defined as a source free annulus surrounding and concentric with each source region
file, to account for the variation of diffuse emission, and to
minimise effects related to the spatial variation of the CCD response. 

The source spectra for the two nuclei were created using the CIAO tool {\em psextract} and fitted in XSPEC (v11.3.1). Due to the low number of counts in this observation the Cash statistic \citep{Cash_79} was used in preference to \chisq\ when modelling the data. Both sources were well fitted with an absorbed thermal model plus an absorbed power law, the absorption component was fixed at the value out of our Galaxy (1.68 $\times 10 ^{20} $ atom cm$^{-2}$). The data were restricted to 0.3$-$6.0 keV, as energies below this have calibration uncertainties, and the spectra presented here do not have significant source flux above 6.0 keV. The parameters from these best fit models can be seen in Table \ref{tab:mkn266_ps}; where columns 2 and 3 give the right ascension and declination (J2000), column 4 the count rate, column 5 the source significance, column 6 the Galactic value of \NH, column 7 $kT$, column 8 metallicity, column 9 power law photon index ($\Gamma$) and columns 10 and 11 give the observed and intrinsic (i.e. corrected for absorption) luminosities. 

\begin{table*}
\centering
\caption[]{Sources detected in the 0.3$-$6.0 keV band within the Mkn 266 with a summary of point source spectral fits. Errors on the spectral
fits parameters are given as 1\( \sigma \) for 1 interesting parameter
from XSPEC. An F denotes that the value has been frozen.
\label{tab:mkn266_ps}}
\begin{tabular}{c@{}c@{}c@{}c@{}c@{}c@{}cccc@{}c@{}}
\noalign{\smallskip}
\hline
Source   & RA & Dec. & Count Rate & Sig. & \multicolumn{4}{c}{Spectral Fit     } & \multicolumn{2}{c}{Luminosity (0.3$-$6.0 keV)   }  \\
       &  & &  & & \NH & kT & Z & $\Gamma$  
       & \multicolumn{2}{c}{(10\(^{41} \) erg s\(^{-1} \)) }    \\ 
       & & &($\times 10^{-2}$ count s$^{-1}$) & ($\sigma$) & ($\times 10 ^{20} $ atom cm$^{-2}$) & (keV) & (\Zsol) & & Observed & Intrinsic   \\      
  \\ \hline

A  & 13:38:17.84 & +48:16:41.2 & 1.90$\pm$0.11 & 41.8 & 1.68 F & 0.80$^{+0.07}_{-0.11}$ & 0.59$^{+1.50}_{-0.24}$ & 0.02$^{+0.25}_{-0.21}$ & 3.41 $\pm 0.17$ & 3.47 $\pm 0.17$ \\
B  & 13:38:17.36 & +48:16:32.5 & 0.90$\pm$0.08 & 18.7 & 1.68 F & 0.88$^{+0.08}_{-0.10}$ & 0.30 F                 & 1.13$^{+0.49}_{-0.64}$ & 0.98 $\pm 0.07$ & 1.04 $\pm 0.07$ \\ 

\noalign{\smallskip}
\hline
\end{tabular}
\end{table*}

From this table it can be seen that source A, the LINER, exhibits a remarkably flat power law of 0.02$^{+0.25}_{-0.21}$. It is likely that part of the contribution to this component arises from unresolved point sources but the very flat slope of this fit suggests that some of the hard component is either heavily absorbed or dominated by reflection \citep{Levenson_04,Matt_00}. By including an absorber for the power law component we can allow for this extra absorption, however, due to the low number of counts, when including this component the fit becomes unconstrained.

It is likely that the cause of this heavy obscuration is due to the strong starburst in this region. An observation by \citet{Sanders_86} found that Mkn 266 is rich in CO, indicating that this system contains a massive, warm reservoir of molecular gas to fuel this starburst, and the dynamic conditions that concentrate this material also serve to further obscure the LINER. The MEKAL component of the fit to this source arises from this starburst, which contributes  5.7$\times$10$^{40}$\ergps\ to the total intrinsic luminosity of source A. 

Source B, the southern nucleus, emits at a temperature of 0.88 keV and has a power law slope of 1.13. This thermal component is likely to arise from the diffuse emission surrounding the nucleus, and the value of $\Gamma$ is consistent with the interpretation of this nucleus being a Seyfert 2 \citep{Cappi_05}. In previous X-ray surveys it has been found that up to $\sim$75\% of Seyfert 2 objects are heavily obscured with \NH\ $\ge$10$^{23}$atom cm$^{-2}$ \citep{Risaliti_99}. With the data we have from this observation the two component model describes the spectrum well, and, due to the limited number of counts, a more complex model with an additional absorption component would over-fit the data. In addition to this, \citet{Cappi_05} have found from a recent survey of nearby Seyfert galaxies, that these sources possess the entire range of \NH\ from 10$^{20}$atom cm$^{-2}$ to 10$^{24}$atom cm$^{-2}$, fairly continuously. Indicating that, although the southern nucleus could be heavily obscured, it is also possible that the two component model described here is good indication of the properties of this source. 
\subsubsection{Diffuse Emission Spatial and Spectral Analysis}
\label{mkn:sec_source_dif}

From the `true colour' image of Mkn 266 it is clear that the galaxy contains significant amounts of diffuse gas, in both the northern feature and also surrounding the central galaxy. To investigate the nature of this diffuse emission, spectra were extracted from these two separate regions using the CIAO tool {\em acisspec}, and fitted in XSPEC. Once again, due to the low number of counts, the Cash statistic was used when modelling the data.  

\begin{figure}
  \hspace{0.1\linewidth}
  \includegraphics[width=0.8\linewidth]{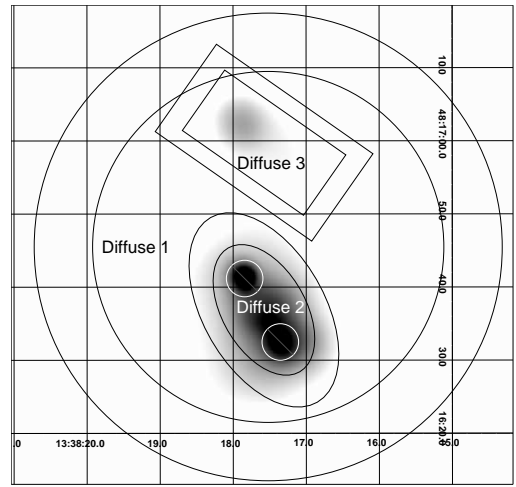}
  \hspace{0.1cm}
  \caption{Adaptively smoothed 0.3$-$8.0\,keV X-ray data from {\em Chandra} ACIS-S of Mkn 266, with extraction and background region for the diffuse gas shown. The background region files are selected to be source free annuli, surrounding and concentric with each extraction region file.}
  \label{fig:mkn_diff}
\end{figure}

The diffuse gas contained within the galaxy is not well described by a single temperature fit, and is better modelled with two temperature components. It is likely that the hotter gas contributing to this model arises from the enhanced emission seen between the two nuclei. To investigate this, the extraction region was divided into two separate parts; one, smaller region, to probe the impact area, where it is likely that the two discs have collided, and a second, larger region, covering the rest of the galaxy, but which excludes the inner impact region and the northern diffuse region. These, along with the extraction region for the northern diffuse emission and the associated background regions, are shown in Figure \ref{fig:mkn_diff}. The two separate regions, diffuse 1 and diffuse 2, are both well described with single component MEKAL fits, exhibiting the same temperatures that were fitted in the two temperature model. The northern region is also well described with a single component MEKAL fit. The parameters from the best fit models for these regions, along with their X-ray luminosities, are shown in Table \ref{tab:mkn_diff}.

\begin{table*}
\centering
\caption[]{Summary of the spectral fits of the diffuse gas regions in Mkn 266. Errors on the spectral
fits parameters are given as 1\( \sigma \) for 1 interesting parameter
from XSPEC. An F denotes that the value has been frozen.
\label{tab:mkn_diff}}
\begin{tabular}{cccccc}
\noalign{\smallskip}
\hline
Diffuse Region  & \NH 					& kT  	& Z 		& \multicolumn{2}{c}{Luminosity (0.3$-$6.0 keV)}  \\
       		& ($\times 10 ^{20} $ atom cm$^{-2}$) 	& (keV) & (\Zsol)	& \multicolumn{2}{c}{(10$^{40}$ erg s$^{-1}$)} \\
		&					&	&		& Observed & Intrinsic   \\      
  \hline

Diffuse 1  	& 1.68 F 				& 0.52$^{+0.06}_{-0.06}$ & 0.23$^{+0.05}_{-0.03}$ 	&  9.88 $\pm 0.50$ & 10.92 $\pm 0.50$ \\
Diffuse 2	& 1.68 F 				& 1.07$^{+0.06}_{-0.09}$ & 0.22$^{+0.15}_{-0.08}$ 	&  8.10	$\pm 0.49$ & 9.83  $\pm 0.53$ \\
Diffuse 3  	& 1.68 F 				& 0.30$^{+0.02}_{-0.03}$ & 0.30F	 		&  6.10 $\pm 0.49$ & 6.90  $\pm 0.55$ \\

\noalign{\smallskip}
\hline
\end{tabular}
\end{table*}

From this table it can be seen that the diffuse gas within the galaxy system, diffuse 1, exhibits a temperature of 0.52 keV, a fairly typical temperature for gas in interacting galaxies (RP98). The impact region, diffuse 2, has a higher temperature of 1.07 keV. This is, within errors, the same temperature attained from the higher temperature contribution from the two component MEKAL fit of the combined spectrum for diffuse 1 and diffuse 2, indicating that the higher temperature from this fit was a consequence of the gas arising from this impact area. 

The northern emission appears spectrally distinct from the surrounding diffuse gas and has been found to emit at a cooler temperature of 0.30 keV. With this \CHANDRA\ data it has also been found that this region has a much lower luminosity than the one derived in \citet{Kollatschny_98}, who report on a \ROSAT\ HRI observation of the galaxy and find the X-ray luminosity of this region to be 3.1$\times$10$^{41}$ \ergps, as opposed to  6.9$\times$10$^{40}$ \ergps, as reported in this work. However, in \citet{Kollatschny_98} this X-ray luminosity is derived from assuming that the relative number of HRI counts is directly proportional to their share of the integrated X-ray flux, with the total luminosity of the system being derived from the \ROSAT\ PSPC observation of the system and consequently has large uncertainties associated with it.

\subsubsection{Nature of the Northern Emission}
\label{sec:mkn266_northern}

The origin of the diffuse emission to the north of the system is still the subject of debate. There have been a number of suggestions as to how this feature has been formed. From HRI and PSPC observations with \ROSAT, \citet{Wang_97} suggest that it is an outflow, driven by the mechanical energy of the supernovae and stellar winds in the starburst. A subsequent paper \citep{Kollatschny_98}, investigating both the HRI observation and optical B,V,R-images of Mkn 266, argued that the scenario proposed by \citet{Wang_97} is unlikely, due to both the high luminosity of the `jet' and also its non-radial geometry. They instead suggest that the mechanism that gives rise to the `jet' is excitation by hot post-shock gas, although they do not conclude where the energy to power this feature would arise from.

\begin{figure}
  \includegraphics[width=0.95\linewidth]{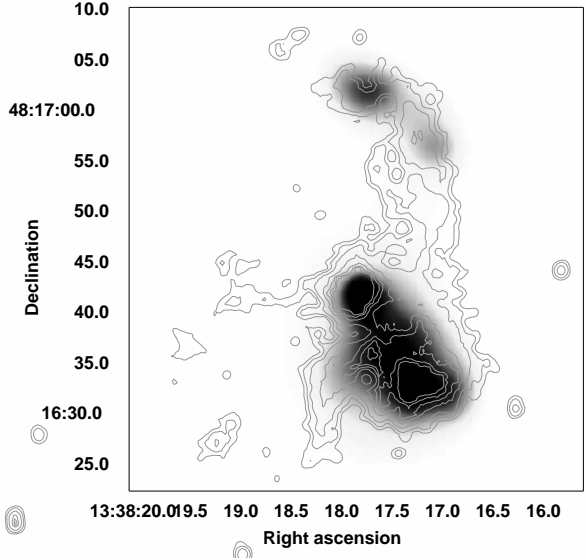}
  \hspace{0.1cm}
  \caption{Adaptively smoothed 0.3$-$8.0\,keV X-ray data from {\em Chandra} ACIS-S of Mkn 266. \Halpha\ contours from \citet{Ishigaki_00} are overlaid. }
  \label{fig:mkn_266_halpha}
\end{figure}

A tridimensional spectrophotometric study by \citet{Ishigaki_00} investigated the \Halpha, [O~\textsc{iii}] and [S~\textsc{ii}] emission-lines within Mkn 266. Figure \ref{fig:mkn_266_halpha} shows the \Halpha\ contours from this observation, overlaid on the full band, adaptively smoothed, \CHANDRA\ X-ray emission. As can be seen, the enhanced regions of \Halpha\ in the northern region are coincident with the two X-ray regions detected with {\em wavdetect} (see Figure \ref{fig:mkn_reg}). Some caution should be exercised when interpreting this plot due to the uncertainty in the astrometry, which we conservatively estimate to be 2\arcs\ (1.1 kpc). Even so, the correlation between the X-ray and \Halpha\ is striking, suggesting that this region to the north of the system is a site of star formation, possibly a tidal arm that has been stripped from the southern progenitor during the merger process. This interpretation is strengthened by the fact that both the {\em HST} (Figure \ref{fig:mkn_rgb}) and \Halpha\ emission seem to connect with the dust lanes around the southern nucleus \citep{Ishigaki_00}.

As shown in the previous section, the luminosity arising from the diffuse gas within the northern feature is 6.9$\times$10$^{40}$ \ergps. Given that our spectral fit to the data indicates that this emission arises from a thermal component, and both the {\em HST} and \Halpha\ emission indicate that there is star formation within this region, the origin of these X-rays could come from supernovae. Making some assumptions about the geometry of the emitting region of this northern feature the thermal energy of the gas can be derived. First the volume, $V$, of the diffuse northern emission has been assumed to be an ellipsoid, with symmetry about the longer axis. The fitted emission measure is equal to $\eta n_\mathrm{e}^{2}V$
and can be used to infer the mean particle number density $n_\mathrm{e}$, with
the filling factor, $\eta$, assumed to be 1. This factor represents the fraction of volume filled by the emitting gas. Although we have assumed this to be 1 in our calculations, there is evidence from hydrodynamical simulations to suggest this value could be $\le$2 per cent \citep{Strickland_00b}. The mean electron density is then used to derive the total gas mass $M_\mathrm{gas}$, which then leads to the thermal energy $E_\mathrm{th}$ of the hot gas. From these assumptions $E_\mathrm{th}$ has been calculated to be between 6.72$\times 10^{54}$ \erg\ (for $\eta$=0.02) and 3.36$\times 10^{56}$ \erg\ (for $\eta$=1).

By deriving the supernova rate, \rSN, for this region, the energy arising from supernovae can be calculated. \citet{Davies_00} calculate \rSN\ for the LINER in Mkn 266, using the relation derived in \citet{Condon_90} 
\begin{equation}
L_{\mathrm{NT}}[\mathrm{WHz^{-1}}]\sim1.3\times 10^{23}(\upsilon[\mathrm{GHz}])^{-0.8}r_{\mathrm{SN}}[\mathrm{yr^{-1}}],
\label{eq:sn}
\end{equation}
and the 20 cm radio continuum \citep{Mazzarella_88}. Where $L_{\mathrm{NT}}$ is the calculated power and $\upsilon$ is the frequency of the observation. From this they calculate \rSN\ to be 0.45 SNyr$^{-1}$. 

For the northern diffuse emission, \citet{Mazzarella_88} find a power of 8.9$\times 10^{21}$WHz$^{-1}$, using this value and the above equation we find \rSN\ =0.05 SNyr$^{-1}$ for this region. From \citet{Mattila_01} it has been found that it is more appropriate to use \LFIR\ to calculate \rSN\ in starburst galaxies but, as we only have \LFIR\ for the whole system, we would greatly overestimate the supernova rate, and consequently, we will use the rate of 0.05 SNyr$^{-1}$ we have derived from the 20 cm continuum data.

From \citet{Davies_00}, models of the star formation timescales have been derived for the LINER. And, although these models give a fairly poorly constrained age (20-500 Myr), by making the assumption that this is a proxy for the timescales of star formation taking place in the northern diffuse emission, the number of supernovae formed in this time can be estimated. Given that we also make the assumption that a massive star takes $\sim 1\times$10$^{7}$ yr to evolve into a supernova, we calculate the number of SNe formed in this time to be $5\times10^5 - 2.5\times10^7$. If each supernova releases $\sim1\times 10^{51}$ \erg, the total thermal energy available will be between $\sim5\times 10^{56}$ \erg\ to $\sim2.5\times 10^{58}$ \erg, thus demonstrating that the observed luminosity in the northern diffuse emission could arise from the contribution of star formation alone. 

The reason that both \citet{Wang_97} and \citet{Kollatschny_98} prefer an outflow scenario to that of tidal stripping to explain this northern feature is due to the presence of optical emission in a highly excited state within this feature. These line ratios, coupled with the electron temperature calculated for this region, cannot be explained by thermal collisional ionisation, and are likely to arise through either photoionisation or shock excitation. \citet{Wang_97} and \citet{Kollatschny_98} both suggest that these shocks arise from a `jet', outflowing from the central galaxy. However, \citet{Ishigaki_00} propose that the northern emission is photoionised by radiation from the Seyfert nucleus and therefore, excitation by shocks is not required to explain the high excitation emission lines that are observed. 

The present study adds further evidence to this debate. With the higher resolution provided by \CHANDRA, the structure of the northern emission can now be resolved. From this it can be seen that the northern feature is curved, and the X-ray emission appears to be `clumpy', with the two most intense areas of X-ray emission coincident with the \Halpha\ emission (see Figure \ref{fig:mkn_266_halpha}). These are not features which indicate a superwind. Furthermore, the morphology of this region traces that of the optical emission, which is highly suggestive that these two features are connected. Consequently, we propose that the most likely source of this X-ray emission is star formation taking place within a tidal arm that has been stripped out from one of the progenitors during the merger of the two galaxies.

\subsubsection{The South East Extension}

To the south east of the central galaxy region there is some suggestion of X-ray extension, which appears to be coincident with filaments seen in the {\em HST} image, just beginning to break out of the galactic disc (Figure \ref{fig:mkn_rgb}). Given the low count statistics in this observation we cannot extract spectra for this region alone. But, from the morphology of the X-ray emission, coupled with the \Halpha\ emission, there is some indication that this region is a site of star formation just on the point of break-out from the gaseous envelope of the central system. This scenario is preferred to the one suggested in the case of the northern emission as there is additional evidence of an outflow feature from the starburst region around the LINER \citep{Ishigaki_00}. This outflow could be beginning to sweep the dust out of the galaxy, indicating that this system is in the stage just prior to the outbreak of large-scale galactic winds throughout the system. Of course, alternatively, this feature could be a small scale version of the northern emission, particularly given its coincidence with optical emission seen with the {\em HST}. However, without spectral information to investigate the nature of this object, neither scenario can be ruled out.

\subsection{Arp 222}
\label{sec:arp222}
\subsubsection{Overall X-ray Structure}

\begin{figure*}
  \begin{minipage}{0.58\linewidth}
  \vspace{0.7cm}
  \includegraphics[width=\linewidth]{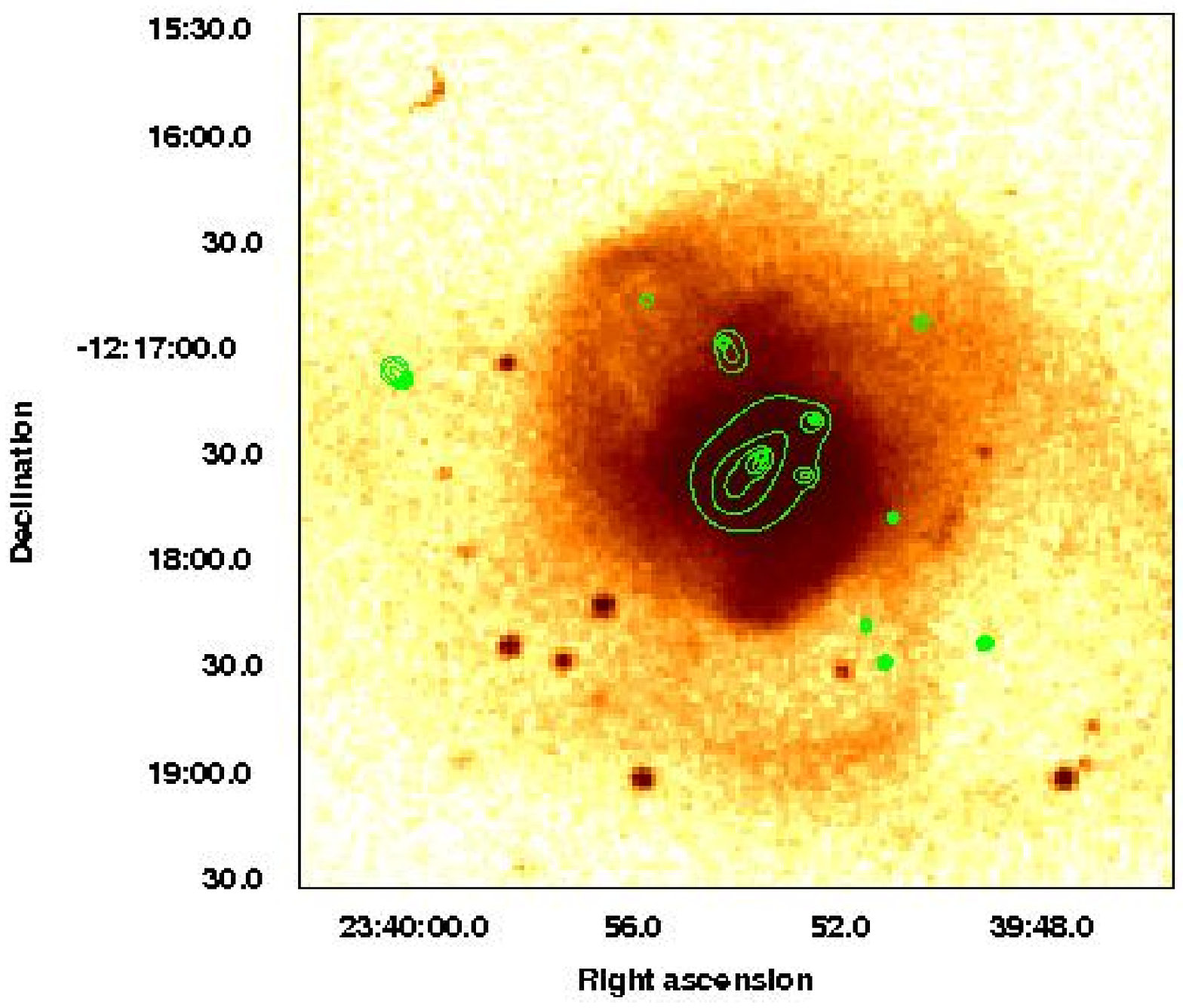}

  \end{minipage}\vspace{0.03\linewidth}
  \begin{minipage}{0.385\linewidth}

  \includegraphics[width=\linewidth]{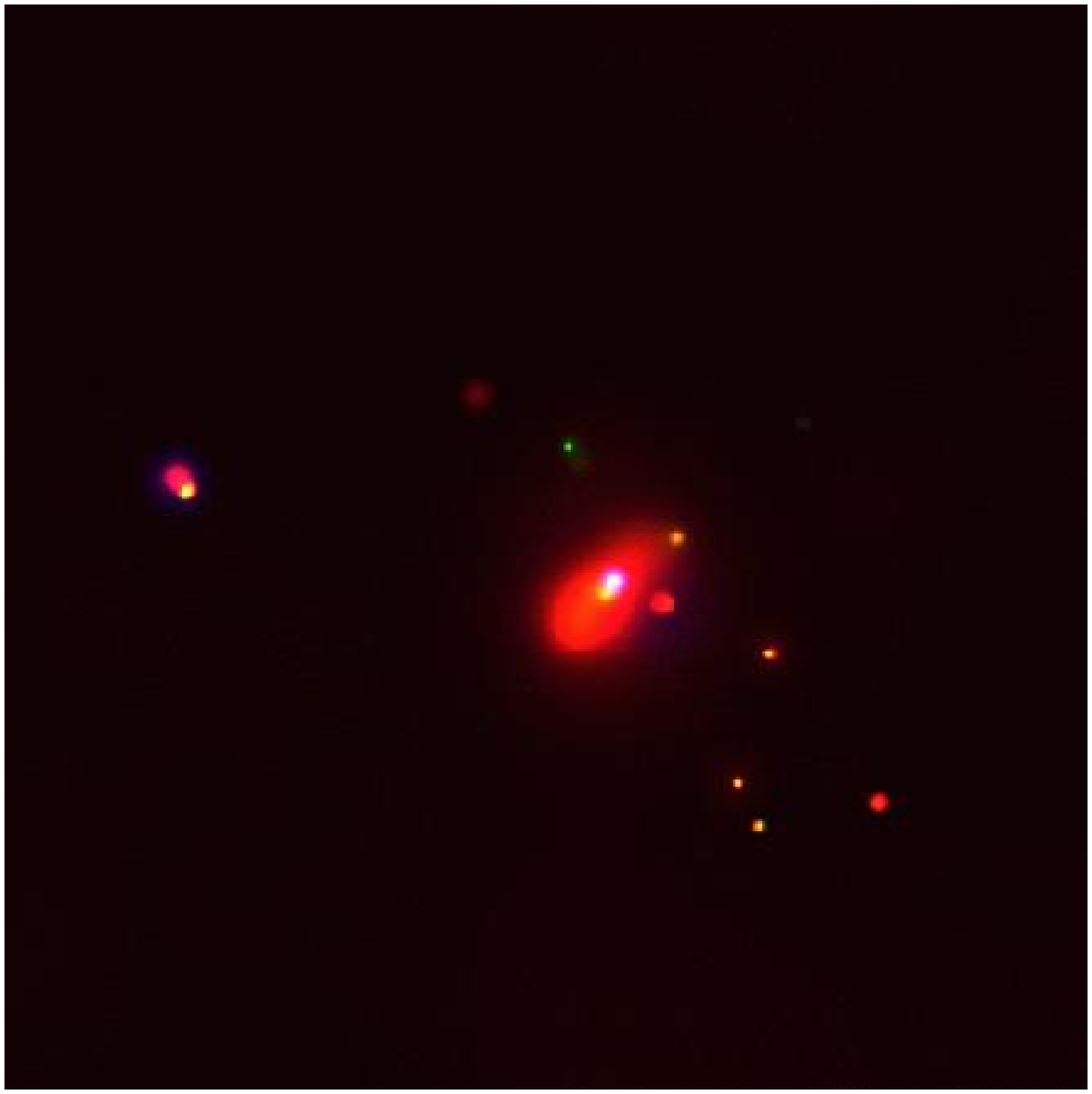}

  \end{minipage}
  \caption{Left, contours of adaptively smoothed 0.3$-$8.0\,keV X-ray data from {\em Chandra} ACIS-S overlaid on an optical image from the UK Schmidt telescope. Right shows the `true colour' image of Arp 222. Red corresponds to 0.3$-$0.8 keV, green to 0.8$-$2.0 keV and blue to 2.0$-$8.0 keV.
  \label{fig:222_gal}}
\end{figure*}

A full band adaptively smoothed \CHANDRA\ image of Arp 222 was produced using the same tools and techniques as described in section \ref{mkn:sec_struc}. In Figure \ref{fig:222_gal} both the optical image from the UK Schmidt telescope, with the smoothed full band X-ray contours overlaid (left), and the `true colour' image (right) of Arp 222 is shown. The true colour image was produced using the same methods as described in section \ref{mkn:sec_struc}.

From these images it can be seen that there is some diffuse gas at the centre of the galaxy, but this emission does not extend out to the optical confines of the system. In addition to the diffuse gas, a number of point sources are found through out the galaxy. From the `true colour' image, the white appearance of the nucleus indicates that it emits in all three energy bands, whilst the other point sources can be seen to be softer.

\subsubsection{Spatial and Spectral Analysis}

\begin{figure}
\hspace{0.1\linewidth}
  \includegraphics[width=0.8\linewidth]{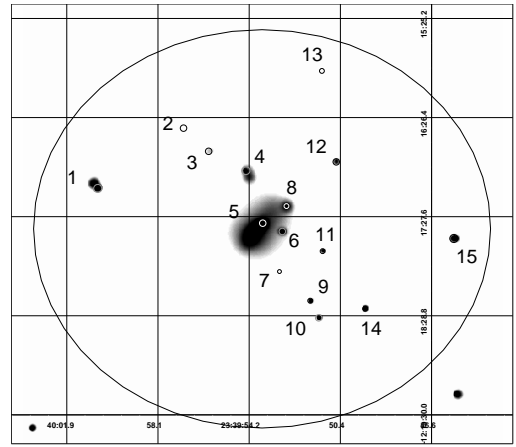}
  \hspace{0.1cm}
  \caption{Adaptively smoothed 0.3$-$8.0\,keV X-ray data from {\em Chandra} ACIS-S of Arp 222, with point sources detected by {\em wavdetect} indicated. The D$_{25}$ ellipse is also shown.}
  \label{fig:arp222_ps}
\end{figure}

X-ray point sources were searched for using the CIAO tool {\em wavdetect}. This was run on the full band image, over the 2, 4, 8, 16 pixel wavelet scales (where pixel width is 0.5\arcs), with a significance threshold of 2.8$\times$10$^{-6}$, which corresponds to one spurious source over a 600 $\times$ 600 pixel grid, the size of our image. 24 sources were detected in this full band range, these were then limited to those that lie within the $D_{25}$ ellipse of the galaxy, reducing the number to 15 detected sources. These regions, along with the $D_{25}$ ellipse are shown in Figure \ref{fig:arp222_ps}. From using the \CHANDRA\ Deep Field South number counts \citep{Giacconi_01} we estimate 2 to 3 of these sources to be background objects. 

Spectra for the point sources were extracted using the CIAO tool {\em psextract} with region and background region files selected in the same way as described in section \ref{mkn:sec_source_dif}. Of the 15 detected sources, only one, source 5, emitted sufficient counts to be modelled individually, the other 14 sources were fitted simultaneously.  The spectrum of the diffuse gas was extracted with the CIAO tool {\em acisspec}, from a region file centred on source 5, with a radius of 45\arcs. The background region file was defined to be a source free annulus surrounding and concentric with the region file. Once again, due to the low number of counts, the Cash statistic was used in preference to \chisq. Both the combined sources and source 5 were well described by single component, power law models. And, from this combined fit model, the luminosities for each individual source were derived. The diffuse gas was well fitted by a single component MEKAL model. These spectral fits are summarised in Table \ref{tab:arp222}, where column 1 gives the source identification number,  columns 2 and 3 give the right ascension and declination (J2000), column 4 the Galactic value of \NH, column 5 $kT$, column 6 metallicity, column 7 $\Gamma$ and columns 8 and 9 give the observed and intrinsic luminosities. 

\begin{table*}
\centering
\caption[]{A summary of the best fit parameters for the diffuse gas and point sources detected in the 0.3$-$6.0 keV band within Arp 222. Errors on the spectral fits parameters are given as 1\( \sigma \) for 1 interesting parameter
from XSPEC. An F denotes that the value has been frozen.
\label{tab:arp222}}
\begin{tabular}{cccc@{}cccc@{}c@{}}
\noalign{\smallskip}
\hline
Source   & RA & Dec. & \multicolumn{4}{c}{Spectral Fit     } & \multicolumn{2}{c}{Luminosity (0.3$-$6.0 keV)   }  \\
       &  & & \NH & kT & Z &$\Gamma$ 
       & \multicolumn{2}{c}{(10\(^{39} \) erg s\(^{-1} \)) }    \\ 
       & & & ($\times 10 ^{20} $ atom cm$^{-2}$) & (keV) & (\Zsol) &  & Observed & Intrinsic   \\      
  \\ \hline

1  & 23:40:00.61 & -12:17:10.2 & 2.75 F & - & - & 1.72F & 0.99 $\pm 0.17$ & 1.07 $\pm 0.17$ \\
2  & 23:39:57.01 & -12:16:33.0 & 2.75 F & - & - & 1.72F & 0.31 $\pm 0.16$ & 0.32 $\pm 0.16$ \\ 
3  & 23:39:55.95 & -12:16:47.3 & 2.75 F & - & - & 1.72F & 0.19 $\pm 0.07$ & 0.21 $\pm 0.08$ \\
4  & 23:39:54.37 & -12:16:59.1 & 2.75 F & - & - & 1.72F & 0.45 $\pm 0.11$ & 0.48 $\pm 0.12$ \\
5  & 23:39:53.67 & -12:17:31.6 & 2.75 F & - & - & 2.15$^{+0.37}_{-0.36}$ & 1.74 $\pm 0.21$ & 1.97 $\pm 0.24$ \\
6  & 23:39:52.84 & -12:17:36.8 & 2.75 F & - & - & 1.72F & 0.37 $\pm 0.11$ & 0.40 $\pm 0.12$ \\
7  & 23:39:52.97 & -12:18:01.5 & 2.75 F & - & - & 1.72F & 0.20 $\pm 0.09$ & 0.21 $\pm 0.09$ \\
8  & 23:39:52.67 & -12:17:21.1 & 2.75 F & - & - & 1.72F & 0.43 $\pm 0.11$ & 0.46 $\pm 0.12$ \\
9  & 23:39:51.67 & -12:18:19.6 & 2.75 F & - & - & 1.72F & 0.52 $\pm 0.14$ & 0.54 $\pm 0.15$ \\
10 & 23:39:51.31 & -12:18:30.0 & 2.75 F & - & - & 1.72F & 0.35 $\pm 0.11$ & 0.37 $\pm 0.11$ \\
11 & 23:39:51.15 & -12:17:48.8 & 2.75 F & - & - & 1.72F & 0.38 $\pm 0.11$ & 0.42 $\pm 0.13$ \\
12 & 23:39:50.57 & -12:16:53.7 & 2.75 F & - & - & 1.72F & 0.28 $\pm 0.09$ & 0.30 $\pm 0.10$ \\
13 & 23:39:51.18 & -12:15:57.6 & 2.75 F & - & - & 1.72F & 0.30 $\pm 0.12$ & 0.31 $\pm 0.13$ \\
14 & 23:39:49.36 & -12:18:24.4 & 2.75 F & - & - & 1.72F & 0.38 $\pm 0.11$ & 0.40 $\pm 0.12$ \\
15 & 23:39:45.63 & -12:17:41.2 & 2.75 F & - & - & 1.72F & 0.21 $\pm 0.12$ & 0.55 $\pm 0.13$ \\
\\
Diffuse  & 23:39:53.67 & -12:17:31.6 & 2.75 F & 0.60$^{+0.07}_{-0.06}$ & 0.08$^{+0.06}_{-0.02}$ & - & 5.48 $\pm$ 0.28 & 6.52 $\pm$ 0.33 \\

\noalign{\smallskip}
\hline
\end{tabular}
\end{table*}

From this table it can be seen that the best-fit spectral index for the combined sources is $\Gamma$=1.72, a typical value for XRBs in, what can be described as, a low/hard state \citep{Soria_03}. For the central source, source 5, the best fit model has shown that the spectral state is much softer, with $\Gamma$=2.15. This value is consistent with it also being an XRB, but this time in a high/soft state \citep{Colbert_04}  This source, along with source 1, has been found to be a ULX. Given the quiescent nature of the galaxy, and that the time since the last episode of widespread star formation has been shown to be $\sim$1.2 Gyr ago \citep{Georgakakis_00}, this indicates that these sources are likely to arise from low mass X-ray binaries (LMXRBs), not HMXRBs as has been seen in younger merger systems within this sample.

The amount of diffuse gas within the system is smaller than in mature elliptical galaxies, and, as mentioned previously, does not extend out to the optical confines of the system. The temperature of the gas, 0.6 keV, has been found to be comparable to that of other interacting systems, but the diffuse gas exhibits a lower X-ray luminosity (6.52$\times 10^{39}$ \ergps) than the other systems within this sample. The low gas content of Arp 222 is further seen from CO observations of the galaxy \citep{Crabtree_94}. From this \CHANDRA\ observation it can be seen that this system is X-ray faint and does not currently resemble a mature elliptical galaxy, although may at a greater dynamical age.

\section{The Evolution of Merging Galaxies}
\label{sec_evol}

\subsection{Multi-wavelength Evolution Sequence}

To gain a greater understanding of the merger process of galaxy pairs, in addition to the X-ray luminosity of each system, B-band, K-band and FIR luminosities have been obtained to study how each of these luminosity diagnostics evolve along the chronological sequence. In the case of systems where the X-ray luminosity has been obtained from the literature, the stated band has been converted into the 0.3$-$6.0 band used in this paper by assuming canonical models of the two X-ray components. For the diffuse gas, a MEKAL model with a gas temperature of 0.5 keV has been assumed, and for the point source population a power law model with a photon index of 1.5 has been adopted.

These luminosities are given in Table \ref{tab:merger:props} where; column 1 gives the system name, column 2 the distance to the object, column 3 the merger age, with nuclear coalescence being defined as 0 Myr, column 4 the total (0.3$-$6.0 keV) intrinsic X-ray luminosity from the \CHANDRA\ observation, column 5 gives the percentage of luminosity arising from the diffuse gas (\%{\ensuremath{L_{\mathrm{diff}}}}), column 6 \LFIR, column 7 \LB\ and column 8 \LK.

\begin{table*}

\centering

\caption[]{The fundamental properties of all the systems within this sample. Columns are explained in the text.
\label{tab:merger:props}}
\begin{tabular}{cccccccc}
\noalign{\smallskip}
\hline

Galaxy 		& Distance	& Merger Age	& \LX			& \%L$_{diff}$	& Log \LFIR			& Log \LB 	& Log \LK \\
System 		& 		&		& (0.3$-$6.0 keV)	&		& 				&		&	    \\
       		& (Mpc)		& (Myr)		& ($\times10^{40}$\ergps) & 		& (\ergps)   		& (\ergps)	& (\ergps)    \\
 
\noalign{\smallskip}
\hline

Arp 270		& 28		& -650 		& 2.95			&	28	&43.63				&43.72		&43.37	    \\
The Mice	& 88		& -500 		& 6.01			&	31	&44.12				&44.07		&44.04	    \\
The Antennae	& 19		& -400 		& 6.97			&	54	&43.95				&43.98		&44.64	    \\
Mkn 266		& 115		& -300 		& 73.18			& 	41	&44.75				&44.30		&44.38	    \\
NGC 3256	& 56		& -200 		& 87.70			&	80	&45.19				&44.42		&45.22	    \\
Arp 220		& 76		& 0    		& 23.70			&	44	&45.50				&43.97		&44.78	    \\
NGC 7252	& 63		& 1000 		& 6.17			&	31	&44.00				&44.40		&44.61	    \\ 
Arp 222		& 23		& 1200 		& 1.46			&	45	&$\geq$42.37		&43.92		&44.84	    \\
NGC 1700	& 54		& 3000 		& 17.58			&	83	&42.91				&44.32		&45.15	    \\

\noalign{\smallskip}
\hline
\end{tabular}
\end{table*}

The FIR luminosities are calculated using the
expression \citep{Devereux_89}
\begin{equation}
\label{eq:lfir}
\LFIR =3.65 \times 10^5[2.58S_{60 \mu m}+S_{100\mu m}]D^2\Lsol,
\label{equ:lfir}
\end{equation}
with {\em IRAS} 60- and 100-$\mu$m fluxes taken from the {\em IRAS}
Point Source Catalogue \citep{Moshir_90}. The optical (B) luminosities
were calculated as in \citet{Tully_88}
\begin{equation}
\mathrm{log} \LB\, (\Lsol) =12.192-0.4B_\mathrm{T}+2\mathrm{log}D,
\end{equation}
where $B_\mathrm{T}$ is the blue apparent magnitude and $D$ is the
distance in Mpc. Values of blue apparent magnitude were taken from
\citet{Dev_91} (the value for The Mice was taken from NGC 2000.0
\citep{Dreyer_88}). The values of \LK\ are derived using the relation given in \citet{Seigar_05}
\begin{equation}
\mathrm{log} \LK=11.364-0.4K_\mathrm{T}+\mathrm{log}(1+Z)+2\mathrm{log}D,
\end{equation}
where $K_\mathrm{T}$ is the K-band apparent magnitude, $Z$ is the galaxy redshift and $D$ is the distance in Mpc. Apparent K-band magnitudes were taken from the 2MASS survey.

To calculate the value of \%{\ensuremath{L_{\mathrm{diff}}}} for each system, the contribution to the luminosity of the diffuse gas that arises from unresolved low luminosity point sources had to be estimated and removed. This was done in a number of ways; in the case of systems for which we have access to the reduced data (The Mice, Mkn 266 and Arp 222), an additional power law component with a photon index of 1.5 was included in the spectral fit of the diffuse emission. It was then assumed that the luminosity of this component arises from unresolved point sources within the system. Additionally, for Mkn 266, it was assumed that the MEKAL component of the point source fits arises from the diffuse gas surrounding the LINER and Seyfert. In the cases of Arp 270 and NGC 7252, estimates of the contribution of the unresolved point sources to the total diffuse luminosity has already been made in \citet{Brassington_05} and \citet{Nolan_04}, respectively.

For the systems we have taken from the literature, the `Universal Luminosity Function' (ULF), derived by \citet{Grimm_03}, was used to predict the flux from unresolved low luminosity sources. The predicted XRB luminosity function is
\begin{equation}
N(>L)=5.4\mathrm{SFR}(L_{38}^{-0.61}-210^{-0.61}),
\label{equ:sfr}
\end{equation}
where $L_{38}$ = $L/10^{38}$ erg s$^{-1}$ and  the factor 210 arises
from the upper luminosity cut-off of 2.1$\times 10^{40}$ erg
s$^{-1}$. 
%
%
SFR was estimated from the population of brighter point sources detected above the completeness limit from the \CHANDRA\ observations. 
In the case of The Antennae, point sources were detected down to a luminosity of 5$\times10^{37}$erg s$^{-1}$ and therefore no correction was required for this system.

As mentioned previously, one of the main difficulties in compiling an evolutionary sequence such as this, is assigning an age to each of these systems. The way in which these estimates were made is described in section \ref{sec:sample}, but another important point that was not discussed is that when constructing an evolutionary sample, the absolute timescale of the merger process must be considered. From N-body simulations \citep{Mihos_96} it is clear that from the first initial strong interaction between equal-mass mergers, through to their nuclear coalescence, takes $\sim$700 Myr. What is not so well defined is the amount of time that elapses between coalescence and the relaxation of the galaxy into a system resembling a mature elliptical galaxy. Within the sample presented here, this issue has been addressed by selecting a greater dynamical range than has previously been studied. By doing this it is hoped that the transition between young merger remnants to relaxed mature ellipticals can be observed, and therefore a more complete picture of the merger process can be obtained. 

Although these merger systems are being compared to establish a single evolutionary sequence, it should be remembered that nine individual systems, not nine examples of one merger at different stages of its evolution, are being looked at. Consequently, although the trends that have been identified here should be a good indicator of how X-ray emission evolves during the merger process, due to the careful selection criteria outlined in section \ref{sec:sample}, it is likely that other merger pair systems will exhibit different X-ray properties. These variations are likely to arise due to the unique interaction parameters associated with each merger system, as well as the variation of gas content and mass of individual galaxies within each system.

With the multi-wavelength luminosities that have been collected for each system (Table \ref{tab:merger:props}), the evolution of the galaxy properties along the merger process has been investigated. The activity levels, a proxy for star formation normalised by galaxy mass, is indicated by {\ensuremath{L_{\mathrm{FIR}}}}/{\ensuremath{L_{\mathrm{K}}}}. To investigate the variation of {\ensuremath{L_{\mathrm{X}}}}, scaled by galaxy mass for each system, both the B-band and the K-band luminosities, {\ensuremath{L_{\mathrm{X}}}}/{\ensuremath{L_{\mathrm{B}}}} and {\ensuremath{L_{\mathrm{X}}}}/{\ensuremath{L_{\mathrm{K}}}} have been used. Both of these have been used as normalisation values as \LB\ can become greatly enhanced during periods of star formation due to the presence of young stars, indicating that \LK\ is likely to provide a more reliable tracer of galaxy mass. Therefore, by plotting \LX\ against both of these values, how great this effect is can be observed.   

\begin{figure*}
  \includegraphics[width=\linewidth]{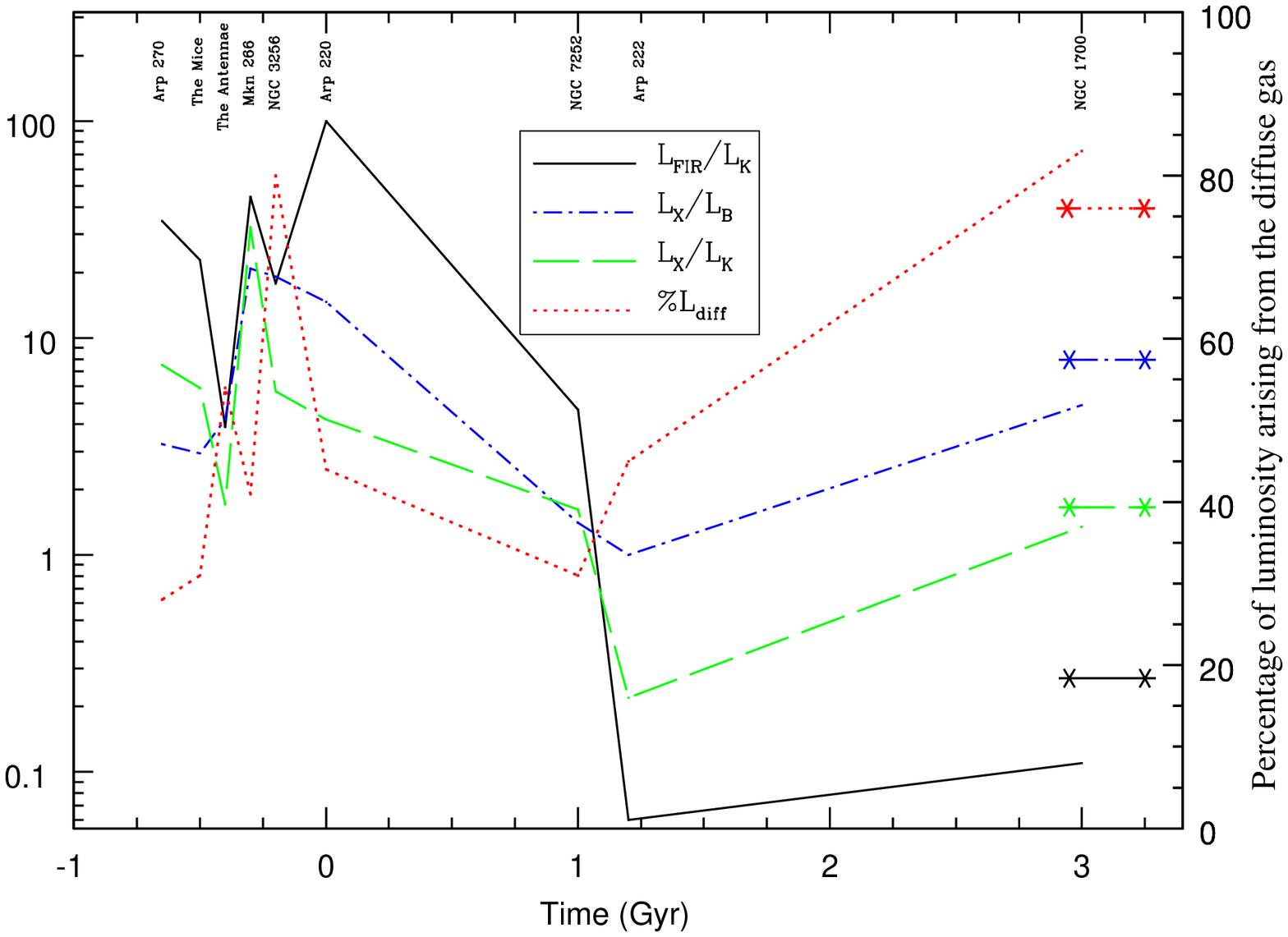}
  \caption {The evolution of X-ray luminosity in merging galaxies. Shown are {\ensuremath{L_{\mathrm{FIR}}}}/{\ensuremath{L_{\mathrm{K}}}} (solid line), {\ensuremath{L_{\mathrm{X}}}}/{\ensuremath{L_{\mathrm{B}}}} (dot-dash line), {\ensuremath{L_{\mathrm{X}}}}/{\ensuremath{L_{\mathrm{K}}}} (dashed line) and \%{\ensuremath{L_{\mathrm{diff}}}}, plotted as a function of merger age, where 0 age is defined to be the time of nuclear coalescence. All luminosity ratios ({\ensuremath{L_{\mathrm{FIR}}}}/{\ensuremath{L_{\mathrm{K}}}}, {\ensuremath{L_{\mathrm{X}}}}/{\ensuremath{L_{\mathrm{B}}}} and {\ensuremath{L_{\mathrm{X}}}}/{\ensuremath{L_{\mathrm{K}}}}) are normalised to the spiral galaxy NGC 2403. \%{\ensuremath{L_{\mathrm{diff}}}} is plotted on a linear scale, shown on the right-hand y-axis of the plot and is an absolute value. \%{\ensuremath{L_{\mathrm{diff}}}} for NGC 2403 is 12\%. The horizontal lines to the right of the plot indicate {\ensuremath{L_{\mathrm{FIR}}}}/{\ensuremath{L_{\mathrm{K}}}}, {\ensuremath{L_{\mathrm{X}}}}/{\ensuremath{L_{\mathrm{B}}}}, {\ensuremath{L_{\mathrm{X}}}}/{\ensuremath{L_{\mathrm{K}}}} and \%{\ensuremath{L_{\mathrm{diff}}}} for NGC 2434, a typical elliptical galaxy.}
  \label{fig:evol}
\end{figure*}

These ratios are shown in Figure \ref{fig:evol}, where, not only the luminosity ratios, but also the percentage of luminosity arising from diffuse gas (\%{\ensuremath{L_{\mathrm{diff}}}}), as a function of merger age have been plotted. {\ensuremath{L_{\mathrm{FIR}}}}/{\ensuremath{L_{\mathrm{K}}}} (solid line), {\ensuremath{L_{\mathrm{X}}}}/{\ensuremath{L_{\mathrm{B}}}} (dot-dash line) and {\ensuremath{L_{\mathrm{X}}}}/{\ensuremath{L_{\mathrm{K}}}} (dashed line) have been normalised to the typical spiral galaxy, NGC 2403. Whilst \%{\ensuremath{L_{\mathrm{diff}}}} (dotted line) is plotted on the right-hand y-axis of the plot, as an absolute value, where \%{\ensuremath{L_{\mathrm{diff}}}} for NGC 2403 is 12\%. The horizontal lines to the right of the plot indicate {\ensuremath{L_{\mathrm{FIR}}}}/{\ensuremath{L_{\mathrm{K}}}}, {\ensuremath{L_{\mathrm{X}}}}/{\ensuremath{L_{\mathrm{B}}}}, {\ensuremath{L_{\mathrm{X}}}}/{\ensuremath{L_{\mathrm{K}}}} and \%{\ensuremath{L_{\mathrm{diff}}}} for NGC 2434, a typical elliptical galaxy \citep{Diehl_06}. 

The system NGC 2403 was selected to represent a typical spiral galaxy. This system was chosen due to its close proximity (D=3.2 Mpc), the fact that it does not have a high level of star-forming activity and that it does not host an AGN \citep{Schlegel_03}. NGC 2434, was selected to represent a typical elliptical galaxy. At a distance of 22 Mpc, this system is close enough to enable the point source population to be disentangled from the hot gas, allowing this diffuse emission to be modelled \citep{Diehl_06}, this systems also exhibits the quiescent levels of star formation normally associated with such galaxies. Table \ref{tab:typical} shows the properties for both of these systems, with columns as in Table \ref{tab:merger:props}.

\begin{table*}

\centering

\caption[]{Properties of the typical spiral (NGC 2403) and the typical elliptical (NGC 2434) used in Figure \ref{fig:evol}. Columns as in Table \ref{tab:merger:props}.
\label{tab:typical}}
\begin{tabular}{ccccccc}
\noalign{\smallskip}
\hline

Galaxy 		& Distance	& \LX			& \%L$_{diff}$	& Log \LFIR	& Log \LB 	& Log \LK \\
System 		& 		&(0.3$-$6.0)			&		& 		&		&	    \\
       		& Mpc		&($\times10^{40}$\ergps)& 		& (\ergps)   		& (\ergps)	& (\ergps)   \\
 
\noalign{\smallskip}
\hline

NGC 2403	& 3		& 0.29			&	12	& 42.21		& 43.22		&43.49	    \\
NGC 2434	& 22		& 4.82			&	76	& 41.96		& 43.54		&44.49	    \\

\noalign{\smallskip}
\hline
\end{tabular}
\end{table*}

Another factor that must be considered when plotting these luminosity ratios is the AGN that is hosted by Mkn 266. This is the only system within the sample that has a confirmed AGN, and as such, it is expected to be significantly more luminous than the other systems. Therefore, the values of {\ensuremath{L_{\mathrm{X}}}}/{\ensuremath{L_{\mathrm{B}}}}, {\ensuremath{L_{\mathrm{X}}}}/{\ensuremath{L_{\mathrm{K}}}} and \%{\ensuremath{L_{\mathrm{diff}}}} have been calculated for Mkn 266, both including and excluding the contribution from the Seyfert. As shown in Table \ref{tab:mkn266_ps}, this nucleus is not a powerful AGN, instead it is close to the Seyfert-LINER borderline, and therefore, the total value of \LX\ only reduces by 14\% if it is excluded, with the value of \%{\ensuremath{L_{\mathrm{diff}}}} rising to 48\%. When the reduced luminosity is used to derive the luminosity ratios for Mkn 266, the change in Figure \ref{fig:evol} is very small, and the overall trends exhibited by these ratios are preserved.

\subsection{Arp 220: the Normalisation of the Universal Luminosity Function}
\label{sec:ulf}

\citet{Grimm_03} considered a number of SFR indicators for each galaxy within their sample. This was done, as there is considerable scatter in the SFR estimates obtained from different indicators. To calculate an `adopted' SFR for each galaxy, indicators that deviated significantly from the other values were disregarded, and the remaining indicators were averaged to give a final value used in subsequent calculations. 

In the present work, the SFR was estimated from the brighter point sources, detected from the \CHANDRA\ observations. To further investigate the normalisation of the ULF, SFR indicators, derived from \LFIR, using the expression \citep{Rosa_02}
\begin{equation}
\mathrm{SFR}_\mathrm{FIR}=4.5 \times 10^{-44}L_\mathrm{FIR}(\mathrm{erg~s}^{-1}),
\end{equation}
have also been used to estimate the luminosity arising from the unresolved point sources. This ULF correction, using the SFR indicator derived from \LFIR, has been calculated for the most active system within our sample, Arp 220. From this correction, a luminosity of 5.05$\times$10$^{41}$erg s$^{-1}$, has been derived for the low luminosity (\LX\ $\le 5\times$10$^{39}$erg s$^{-1}$) point sources. This value is almost two times greater than the {\em total} observed luminosity of Arp 220 (see Table \ref{tab:merger:props}), indicating that using \LFIR\ as a SFR in the ULF must greatly overestimate the point source population for this system. To ensure that the value calculated in equation \ref{equ:lfir} was consistent with other observations, additional \LFIR\ values were obtained (Log \LFIR=45.86, \citet{David_92} and Log \LFIR=45.54, \citet{Liu_95}). Averaging these values gives Log \LFIR\ = 45.73 for Arp 220, similar to the value given in Table \ref{tab:merger:props}.  

To investigate this further, we compare the observed population of brighter sources with that predicted using the ULF. From the \CHANDRA\ observation, 4 sources with \LX $\ge 5\times$10$^{39}$erg s$^{-1}$ are detected in Arp 220. Using the SFR derived from \LFIR\ in equation \ref{equ:sfr}, the bright point source population is estimated to be ten times larger, with 41 discrete sources predicted. This demonstrates that, in the case of Arp 220, when using \LFIR\ as the SFR indicator, the ULF greatly overestimates the whole point source population. {\ensuremath{L_{\ensuremath{\mathrm{H}{\alpha}}}}} values for this galaxy were also used to calculate the SFR of the system \citep{Young_96,Colina_04} but, due to the very dusty nature of this object, are subject to large extinction effects, leading to greatly reduced SFRs with values as little as 0.55, which, given the merger status of this system, is unlikely to be a true measure of the SFR.

\begin{figure}[h]
  \hspace{0.8cm}
  \includegraphics[width=0.8\linewidth]{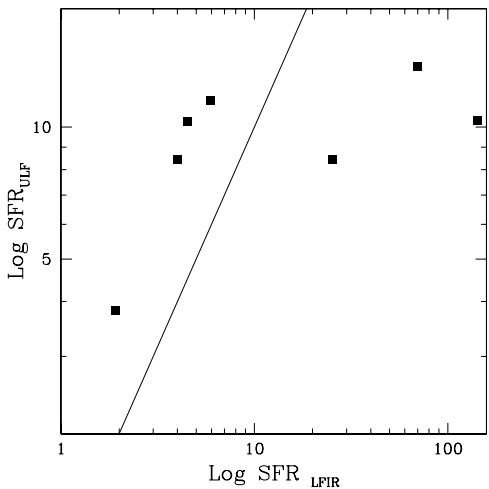}
  \caption{A plot of the SFR indicated by \LFIR\ and the SFR derived from equation \ref{equ:sfr}, with the solid line indicating unity. Arp 222 and NGC 1700 have not been included in this plot as their point source populations are likely to be dominated by LMXRBs.}
  \label{fig:sfr}
\end{figure}

To investigate the reliability of using the \LFIR\ SFR indicator to normalise the ULF, the detected point source populations (with \LX\ $\le 2.1\times10^{40}$ erg s$^{-1}$) from the \CHANDRA\ observations were used to derive SFRs from equation \ref{equ:sfr} for a number of galaxies in our sample; these values are plotted against the \LFIR\ derived SFR values in Figure \ref{fig:sfr}. Only seven of the nine systems have been included in this figure, as the ULF is only a measure of the HMXRB population, and, it is likely that the stellar populations from the two older merger-remnants are dominated by LMXRBs.
From Figure \ref{fig:sfr}, it can be seen that at high levels of the
\LFIR\ SFR indicator, the SFR derived from equation \ref{equ:sfr} is
much smaller, indicating that, in very active star-forming systems,
the prediction of the point source population, when using the \LFIR\
SFR value, does not represent the observed X-ray population well. 

This relationship between \LX\ and SFR was also investigated in
\citet{Gilfanov_04b}, where active systems in the HDF-N  were used to
probe the ULF at high values of SFR. From this work it was found that,
for higher levels of star formation, the observed X-ray luminosities
from these galaxies do follow the \LX-SFR relation. However, the
calculated SFR for all of these systems were derived from 1.4 GHz, {\ensuremath{L_\nu}} SFR
indicators only, not \LFIR\ SFR indicators. This suggests that the discrepancy
between \LFIR\ SFR, and the observed values derived from the
\CHANDRA\ observations, are a consequence of the normalisation
indicator chosen, and do not arise from non-linearity of the ULF at high values of SFR.

There are a number of reasons why the SFRs inferred from the X-ray observations and \LFIR\ values might differ so greatly for systems in the most active phases. It could simply be that, for dusty, active systems such as these, \LFIR\ is a poor tracer of SFR, becoming greatly enhanced due to the large amounts of gas residing in the galaxy. This would indicate that it is the \LFIR\ SFR indicator, not the SFR from the ULF, that is incorrect. However, this explanation is unlikely, as studies have shown that, whilst {\ensuremath{L_{\ensuremath{\mathrm{H}{\alpha}}}}} and {\ensuremath{L_{\mathrm{UV}}}} are susceptible to extinction caused by the presence of dust in star-forming regions, \LFIR\ does not vary with gas content, and scales well with both extinction corrected {\ensuremath{L_{\mathrm{UV}}}} and radio continuum emission in dusty starbursts \citep{Buat_96,Buat_92}.

Alternatively, the discrepancy between the predicted and observed values could be a consequence of the age of the stellar population. If the starburst has only recently taken place, it is plausible that, whilst \LFIR\ has had enough time to become massively enhanced, the X-ray point source population has not yet evolved into HMXRBs. The consequence of this would be that the X-ray luminosity arising from the point source population is lower than the SFR, indicated by \LFIR, would predict. \citet{Wilson_05} recently reported on an optical study of Arp 220 which has detected two populations of young massive star clusters. The age of the youngest of these populations has been found to be 5$-$10 Myr, indicating that enough time has elapsed since the last starburst for HMXRBs to form ($\sim1-$10 Myr). It is therefore unlikely that a lag between \LFIR\ and \LX\ is the reason for the difference in the SFR values shown in Figure \ref{fig:sfr}.

Another reason that the SFR derived from equation \ref{equ:sfr} is lower than expected, when compared to the \LFIR\ SFR, could simply be that the X-ray observations are not detecting all the point sources in very active galaxies due to obscuration, and, those that are being detected have greatly reduced luminosities. To investigate this, the optical obscuration in Arp 220 \citep{Shioya_01} has been converted into an absorbing column density, giving \NH = 2.2$\times 10 ^{22} $ atom cm$^{-2}$, a value which is actually somewhat smaller than the modelled value given in \citet{Clements_02} of 3$\times 10 ^{22} $ atom cm$^{-2}$. This demonstrates that the stated luminosities in \citet{Clements_02} are reliable values, and all the point sources with \LX $\ge 5\times$10$^{39}$erg s$^{-1}$ (the stated lower luminosity point source detection threshold) should have been detected in this system. 

A final explanation for the difference between the predicted and observed SFR indicators for systems close to the point of nuclear coalescence, could arise from a physical change in the birth of stellar systems. These merging galaxies provide extreme examples of violent star formation, and it is therefore possible that the most recent starburst has resulted in the birth of a stellar population with a different Initial Mass Function (IMF). 

The IMF, the distribution of masses with which stars are formed, has long been a contentious issue, with a variety of different models used to explain the observed stellar populations \citep{Larson_86,Larson_98}. It was initially proposed that the IMF was a universal function with a power law form, regardless of formation time or local environment \citep{Salpeter_55}.  But, it has also been argued that the IMF varies, and was more top-heavy at earlier times \citep{Rieke_80}. 

From the most recent observations it is thought that the Salpeter IMF,
with a lower mass flattening between 0.5 \Msol\ and 1.0 \Msol, can be
considered a reasonable approximation for large regions of starburst
galaxies \citep{Elmegreen_05}, with some suggestion of small
variations with environment, with denser and more massive star
clusters producing more massive stars compared to intermediate mass
stars \citep{Elmegreen_04,Shadmehri_04}. This more top-heavy IMF has
the effect of generating greater numbers of supernova remnants, black
holes, and HMXRBs, as well as higher amounts of \LFIR\ per unit mass
of stars formed, therefore leading to an increase in both \LX\ and
\LFIR. However, the
level of enhancement of \LFIR, relative to the increase in HMXRBs formed, is
dependent on the shape of the IMF. 

In the galaxy systems close to the point of nuclear coalescence
presented in this sample,
the X-ray binary population is dominated by HMXRBs, meaning that \LX\
will scale with the number of massive stars produced, whereas \LFIR\
will scale
with the main sequence luminosity of those stars. As the IMF flattens,
\LFIR\ will rise more steeply, a consequence of the strong mass to
luminosity dependence, $M \propto L^{3.5}$. Therefore, when
using \LFIR\ values to derive the SFR for these system, this value will be
overestimated.



If this interpretation is correct, it could explain why there appears to be a deficit of X-ray binaries in the \CHANDRA\ observations of the systems close to the point of coalescence. In actual fact, what is being seen here is an overestimated SFR, derived from the massively enhanced \LFIR, as can be seen in Figure \ref{fig:sfr}.




\section{Discussion: X-ray Evolution}
\label{sec:dis}

The first thing to note from Figure \ref{fig:evol} is that Arp 270, even though an early stage system, is already exhibiting enhanced star formation activity, as measured by \LFIR/\LK\ (see \citet{Read_01}), compared to the typical spiral system, NGC 2403. This activity increases up to the point of nuclear coalescence, after which there is a steady drop in \LFIR/\LK\ until the merger-remnant systems, $\sim$1 Gyr after coalescence. From this point the activity value levels off to that of a typical elliptical system. 

The evolution of the X-ray luminosity is different to that of \LFIR/\LK. There is initially a rise, as seen with the star formation activity, but this peaks $\sim$300 Myr before coalescence takes place, whilst \LFIR/\LK\ is still increasing. From this peak it drops until the young merger-remnants, as is the case for \LFIR/\LK. But, instead of levelling off, the X-ray luminosity once again begins to rise, with the total X-ray luminosity of the 3 Gyr system beginning to resemble that of a mature elliptical galaxy.

Both \LX/\LB\ and \LX/\LK\ broadly exhibit the same variations. One notable exception is the value of \LX/\LB\ for Mkn 266. This relatively small value is due the massive enhancement of \LB\ from this system's AGN \citep{Risaliti_04}, which \LK\ is not susceptible to, therefore indicating that \LK\ is a more reliable tracer of galaxy mass than \LB, and will exhibit less scatter.

The evolution of \%{\ensuremath{L_{\mathrm{diff}}}} also shows an increased value for Arp 270, when compared to the typical spiral system. This value then steadily rises,  up to a point $\sim$200 Myr prior to nuclear coalescence. This trend line then drops until the young merger-remnant systems, once again rising, with the 3 Gyr system exhibiting a similar value of \%{\ensuremath{L_{\mathrm{diff}}}} to that of the typical elliptical, NGC 2434.

\subsection{Previous X-ray Studies}
\label{sec:RP98}

In RP98 it was found that the X-ray luminosity of the sample generally followed the same patterns as the star formation activity, with the peak of \LX/\LB\ being coincident with nuclear coalescence. Also in the RP98 study, the sample of post-merger systems was limited, with a baseline extending out to only 1.5 Gyr after nuclear coalescence. Consequently, no rise in the X-ray luminosity of post-merger systems was observed, because, as shown in the present study, there is a strong suggestion that systems require a much greater relaxation time before they begin to exhibit properties seen in mature elliptical galaxies. This idea is further strengthened by a study of post-merger ellipticals \citep{Osul_01}, where a long term trend ($\sim$10 Gyr) for \LX/\LB\ to increase with dynamical age was found.

The reason that the peak in \LX\ has been found at two different merger ages in these studies could be due to the different observatories that were used. \CHANDRA\ has greater spatial resolution than \ROSAT, and is able to disentangle background galaxies from the target object with much greater accuracy. It is therefore possible that these objects were included in the previous \ROSAT\ work as diffuse features, leading to a different peak value of \LX. However, both samples include the famous merger system, The Antennae, and the system at coalescence, Arp 220, and the X-ray luminosities from both of these systems are comparable between \ROSAT\ and \CHANDRA. Indicating that this discrepancy does not arise from a difference between these two observatories. 

Another difference between RP98 and the current work, that could account for the differing merger ages of peak \LX, is the selection of systems within each study. In our sample we include two systems between The Antennae and Arp 220; Mkn 266 and NGC 3256, in RP98 there is only one, NGC 520. This system is highlighted in RP98 as being X-ray faint, not exhibiting the X-ray properties that one would expect, given its large \LFIR/\LK\ ratio. They suggest that this galaxy does not appear to be on the same evolutionary path as the rest of their sample and, possibly, will not evolve into an elliptical galaxy.

A recent study of NGC 520 with \CHANDRA\ \citep{Read_05}, has indicated that this system comprises one gas-rich and one gas-poor galaxy, not two gas rich spirals as has been selected within this sample. This lack of gas in the second galaxy has resulted in a `half merger' being induced in this system, and hence appears underluminous in X-rays. Therefore, it is likely that the peak X-ray luminosity was missed in the RP98 sample due to the galaxy selection. In the present study, use of the selection criteria outlined in section \ref{sec:sample} has meant that only systems that are gas rich have been included, ensuring that comparable systems in the merger sequence have been selected. Consequently, the systems presented in this sample are more likely to be representative of the typical merger evolution.

\subsection{The X-ray Luminosity Peak and the Impact of Galactic Winds}
\label{sec:galWs}

From the X-ray luminosities of the systems within this sample it has been seen, for the first time, that both \LX/\LK\ and \LX/\LB\ peak $\sim$300 Myr before nuclear coalescence takes place. This result, given that the normalised star formation rate is still increasing at this time, is initially surprising, as one would expect the X-ray luminosity to increase with increasing \LFIR/\LK. However, from looking at the morphology of the hot diffuse gas, it is clear that the systems emitting very high levels of \LFIR/\LK\ are also experiencing outflows from starburst-driven winds. We propose that this relative reduction in X-ray luminosity, for these very active systems, is a consequence of these large scale diffuse outflows.

Starburst-driven winds are responsible for the transport of gas and energy out of star-forming galaxies. The energetics of these galactic winds were investigated by \citet{Strickland_00b}, where hydrodynamical simulations were compared to the observations of the galactic wind of M82. From this work it was shown that the majority of the thermal and kinetic energy of galactic winds is in a hot volume-filling component of the gas, which is very difficult to probe due to its low emissivity, a consequence of the low density of this component.

Within our sample, the system that represents the peak X-ray luminosity is Mkn 266. In section \ref{sec:mkn266} the results from the \CHANDRA\ observation of this system are presented and tentative evidence suggesting that this system is just about to experience galaxy wide diffuse outflows is found. If this interpretation is correct, it seems probable that the lower values of \LX/\LB\ and \LX/\LK, for NGC 3256 and Arp 220, the most active systems in this sample, are a consequence of their extensive galactic winds, which have allowed the density, and hence the emissivity, of the hot gas to drop.

\subsection{Halo Regeneration in Low \LX\ Systems}
\label{sec:halo}

Another interesting result from this survey is the increase in \LX\ in the older merger-remnants. In previous studies (e.g. RP98) post-merger systems have been found to be X-ray faint when compared to elliptical galaxies. In our study the merger sequence has been extended to include a 3 Gyr system, and, by doing so, it has been shown that these underluminous systems appear to increase in \LX\ as they age. Given that these merger-remnants have been shown to be quiescent, this increase in \LX\ is not due to any starburst activity within the system. Coupling this, with the increase in \%{\ensuremath{L_{\mathrm{diff}}}}, indicates that diffuse X-ray gas is being produced, leading to the creation of X-ray haloes, as observed in mature elliptical galaxies. \citet{Osul_01} investigated the relationship between {\ensuremath{L_{\mathrm{X}}}}/{\ensuremath{L_{\mathrm{B}}}} and spectroscopic age in post-merger ellipticals and found that there was a long term trend ($\sim$10 Gyr) for {\ensuremath{L_{\mathrm{X}}}} to increase with time. The mechanism by which the regeneration of hot gas haloes in these galaxies is explained, is one in which an outflowing wind to hydrostatic halo phase is driven by a declining SNIa rate. \citet{Osul_01} argue that a scenario in which gas, driven out during the starburst, infalls onto the
existing halo is not the dominant mechanism in generating X-ray haloes as this mechanism would only take $\sim$1$-$2 Gyr and would therefore not produce the long-term trend they observe. The time baseline from the sample studied in the present paper is not sufficient to allow us to discriminate between these two possibilities.

\subsection{The Behaviour of the X-ray Point Source Population}
\label{sec:compare}

\begin{figure}
  \includegraphics[width=\linewidth]{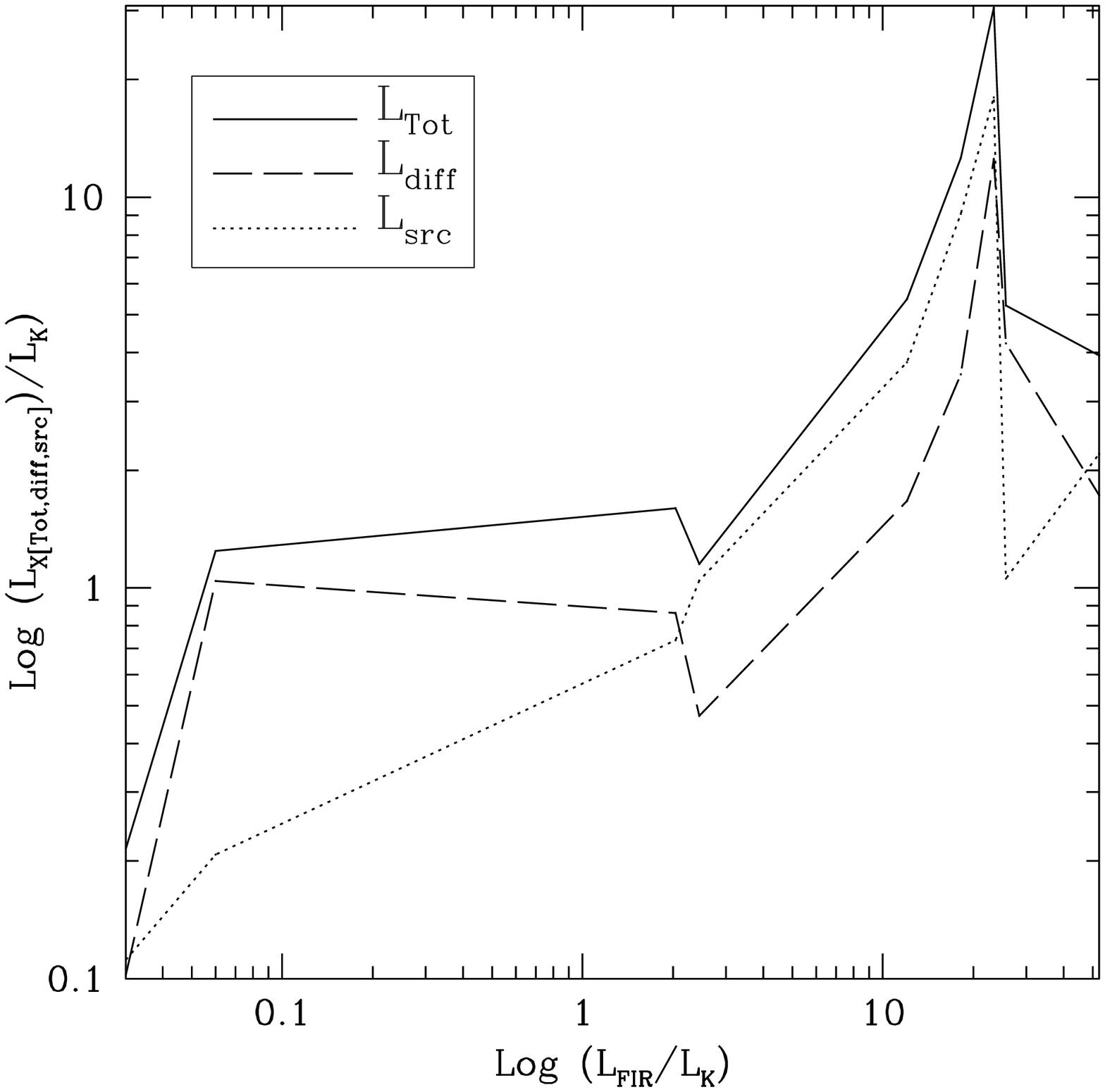}
  \caption{Plot indicating how \LX/\LK\ and the two components that contribute to \LX/\LK\ (the point source contribution (\Lsrc) and the diffuse gas contribution (\Ldiff)) scale with \LFIR/\LK.}
  \label{fig:lx_lps}
\end{figure}

From the \CHANDRA\ observations in this study, the behaviour of the
point source population during the merger process has been
characterised for the first time. In Figure \ref{fig:lx_lps} the total
luminosity, as well as its two components, the luminosity arising from
the point source population (\Lsrc) and the diffuse gas contribution
(\Ldiff), have been normalised by \LK\ and plotted against star
formation activity (\LFIR/\LK).



From this figure it can be seen that the total \LX/\LK, whilst
showing a general trend of increasing with \LFIR/\LK, exhibits a large
amount of scatter, peaking at the merger system Mkn 266, and then dropping at high values of \LFIR/\LK. Decomposing
\LX/\LK\ into separate source and diffuse contributions to
\LX, it can be seen that the scatter arises
primarily from the diffuse component, whilst the drop in
\LX/\LK\ at high \LFIR/\LK, is a consequence of both \Ldiff/\LK\ and
\Lsrc/\LK\ falling steeply. Using the Kendall Rank coefficient, it was found
that the point source component, \Lsrc/\LK, shows a correlation of
2.5$\sigma$ with \LFIR/\LK. Fitting a single power law to these data gives
a logarithmic index of 0.93$\pm$0.28.

However, from visual inspection of Figure \ref{fig:lx_lps}, a single
power law trend does not represent the behaviour adequately.
At high values of \LFIR/\LK, the very active
systems in this sample actually exhibit declining values of
\LX/\LK\ as \LFIR/\LK\ increases. In section
\ref{sec:ulf}, we proposed the possibility that these systems close to
nuclear coalescence have a value of \LFIR\ enhanced relative to that in
less extreme environments, due to a change in the IMF. Under this
hypothesis, the corresponding points in Figure \ref{fig:lx_lps} will have
been shifted strongly to the right, which could account for the
negative slope 
seen in all three curves at high \LFIR/\LK.
The idea that the change in behaviour relates to \LFIR/\LK, rather than
\LX/\LK, is attractive since, given the very different
mechanisms responsible for generating the source and diffuse
components of \LX\ (X-ray binaries and supernova-heated gas
respectively), it is hard to envisage any process which could
simultaneously change the behaviour of both.


\section{Conclusions}
\label{sec:con}

From a sample of nine interacting and merging galaxies, the evolution of X-ray emission, ranging from detached spiral pairs through to merger-remnant systems, has been investigated. As part of this survey, results from the analysis of two \CHANDRA\ observations, Mkn 266 and Arp 222, have also been presented. Here we summarise the results from these \CHANDRA\ observations and then draw the main conclusions from the survey:

\subsection{Mkn 266 and Arp 222}

\begin{itemize}

\item{The luminous merger system, Mkn 266, has been shown to contain two nuclei, with X-ray luminosities of 3.47$\times$10$^{41}$\ergps\ and 1.04$\times$10$^{41}$\ergps. In addition, an area of enhanced X-ray emission has been detected between them. This is coincident with a radio source and is likely a consequence of the collision between the two discs from the progenitors. A region of diffuse emission is detected to the north of the system. It is probable that this arises from a spiral arm that has been stripped out of the system during the merger. To the south east of the nucleus, a region of extended emission has been detected, indicating that this system could be on the verge of large-scale galactic winds breaking out. This system has the highest \LX/\LK\ ratio of any of the galaxies within our sample.} 

\item{ 15 discrete X-ray sources have been detected in Arp 222, two of which are classified as ULXs. This merger-remnant system has been shown to be X-ray faint when compared to both other systems within this sample and the typical elliptical galaxy NGC 2434. The diffuse gas of Arp 222 has been modelled with a temperature of 0.6 keV and, even though optical and CO observations are consistent with those of elliptical galaxies, the X-ray luminosity of Arp 222 does not resemble that of a mature elliptical.}

\end{itemize}

\subsection{The X-ray Evolution of Merging Galaxies}

\begin{itemize}

\item{The most striking result from this work is the time at which \LX/\LB\ and \LX/\LK\ peak. It was previously believed that this was coincident with nuclear coalescence, but here we find the peak $\sim$300 Myr before this coalescence takes place. We suggest that subsequent drop in X-ray emission is a consequence of large-scale diffuse outflows breaking out of the galactic discs, reducing the hot gas density and allowing the escape of energy in kinetic form.}

\item{This study has also demonstrated that, in the systems close to the point of coalescence, \LFIR\ is massively enhanced when compared to the X-ray binary luminosity of these systems. We suggest here that the high level of \LFIR\ result from a change in the IMF in these exceptional starbursts. With the production of more massive stars compared to intermediate mass stars in these galaxies leading to larger values of \LFIR\ per unit mass of stars formed. }

\item{At a time $\sim$1 Gyr after coalescence, the merger-remnants in our sample are X-ray faint when compared to typical X-ray luminosities of mature elliptical galaxies. However, we see evidence that these systems will start to resemble typical elliptical galaxies at a greater dynamical age, given the properties of the 3 Gyr system within our sample. This supports the idea that halo regeneration will take place within low {\ensuremath{L_{\mathrm{X}}}} merger-remnants. We caution that, with only one older, more relaxed, system within our sample, our conclusions on this point are necessarily tentative. To fully understand how young merger-remnants evolve into typical elliptical systems, the period in which this transformation takes place needs to be studied in greater detail.}

\end{itemize}


\section{Acknowledgements}

We thank the \CHANDRA\ X-ray Center (CXC) Data Systems and Science
Data Systems teams for developing the software used for the reduction (SDP) and analysis (CIAO). We would also like to thank the anonymous referee for helpful comments
which improved this paper, and Steve Diehl for providing us with the X-ray information for NGC 1700 and NGC 2403. 

This publication has made use of data products from the Two Micron All Sky Survey, which is a collaboration between The University of Massachusetts and the Infrared Processing and Analysis Center (JPL/ Caltech). Funding is provided by the National Aeronautics and Space Administration and the National Science Foundation.

NJB acknowledges the support of a PPARC studentship.


\label{lastpage}

\bibliographystyle{mn2e}
\bibliography{nicky}

\end{document}